\documentclass[reqno]{amsart}
\usepackage{graphicx}
\usepackage{enumitem}
\usepackage{todonotes}
\usepackage{booktabs}
\usepackage{dsfont}
\usepackage{xcolor}
\usepackage{colortbl}
\usepackage{multirow,array}
\usepackage{amsaddr}
\usepackage[final,color,notref,notcite]{showkeys}

\usepackage[charter]{mathdesign}
\usepackage[scaled=.96,lf]{XCharter}
\usepackage[hypcap = true]{subcaption}
\usepackage{scalerel}

\usepackage{hyperref}
\hypersetup{colorlinks,citecolor=blue}

\usepackage{algorithm}
\usepackage{algorithmic}

\definecolor{labelkey}{rgb}{0,.56,.7}

\DeclareMathOperator*{\bigboxplus}{\scalerel*{\boxplus}{\textstyle\sum}}

\theoremstyle{definition}

\newtheorem{exa}{Example}

\newcommand{\nn}{\nonumber}

\def\dg{\dagger}
\def\df{\overset{\mathrm{df}}{=}}
\newcommand{\ket}[1]{\mathop{|#1\rangle}\nolimits}
\newcommand{\bra}[1]{\mathop{\langle#1|}\nolimits}

\let\bs\boldsymbol

\let\ol\overline
\let\ul\underline

\def\bbN{\mathbb{N}}
\def\bbR{\mathbb{R}}

\newcommand{\n}{\bs{n}}
\newcommand{\bb}{\bs{b}}
\def\bvt{\bs{\vartheta}}
\def\bpsi{\bs{\psi}}

\DeclareSymbolFont{Eulerscripteusm10}{U}{eus}{m}{n}
\SetSymbolFont{Eulerscripteusm10}{bold}{U}{eus}{b}{n}
\DeclareMathSymbol{\euD}{\mathord}{Eulerscripteusm10}{"44}
\DeclareMathSymbol{\rH}{\mathord}{Eulerscripteusm10}{"48}
\DeclareMathSymbol{\rB}{\mathord}{Eulerscripteusm10}{"42}
\DeclareMathSymbol{\rC}{\mathord}{Eulerscripteusm10}{"43}
\DeclareMathSymbol{\rL}{\mathord}{Eulerscripteusm10}{"4C}

\def\H{\mathcal{H}}

\DeclareMathAlphabet{\pazocal}{OMS}{zplm}{m}{n}   

\def\a{\alpha}
\def\vt{\vartheta}

\allowdisplaybreaks

\def\a{\alpha}
\def\b{\beta}
\def\g{\gamma}
\def\d{\delta}

\def\om{\omega}

\def\s{\sigma}

\def\la{\lambda}

\def\vt{\vartheta}

\def\om{\omega}

\begin{document}

\title{Certain properties and applications of shallow bosonic circuits}

\author{Kamil Br\'adler and Hugo Wallner}

\address{ORCA Computing}

\begin{abstract}
We introduce a novel approach to solve optimization problems  on a boson sampling device assisted by classical machine-learning techniques. By virtue of the parity function, we map all measurement patterns, which label the basis spanning an $M$-mode bosonic Hilbert space, to the Hilbert space of~$M$~qubits. As a result, the sampled probability function can be interpreted as a result of sampling a multiqubit circuit. The method is presented on several instances of a QUBO/Ising problem as well as portfolio optimization problems.

Among many demonstrated properties of the parity function is the ability to chart the entire qubit Hilbert space no matter how shallow the initial bosonic circuits is.  In order to show this we link boson sampling circuits to a class of finite Young's lattices (a special poset with the so-called Ferrers diagrams ordered by inclusion), Boolean lattices and the properties of Dyck/staircase paths on integer lattices.  Our results and methods can be applied to a large variety of photonic circuits, including the deep ones of essentially any geometry, but our main focus is on shallow circuits as they are less affected by photon loss and relatively easy to implement in the form of a time-bin interferometer.
\end{abstract}

\email{kamil/hugo@orcacomputing.com}

\date{\today}

\maketitle

\thispagestyle{empty}

\section{Introduction}\label{s:intro}

The origin of boson sampler dates back to one of the first universal proposals for quantum computing for photonics known as the KLM~\cite{knill2001scheme} protocol (after Knill, Laflamme and Milburn). The protocol turn out to be ultimately non-scalable but it was a starting point for a flurry of activities culminating in the development of a truly scalable and fault-tolerant quantum computing paradigm known as measurement-based quantum computing (MBQC)~\cite{raussendorf2001one}. MBQC on a discrete photonic substrate is currently the most developed candidate for a universal, large-scale and fault-tolerant quantum computer from the architectural point of view~\cite{bombin2021interleaving,omkar2021all}.

Even though the KLM proposal does not lead to a scalable quantum computer architecture, its building blocks are  worth of investigating. A central part of KLM is a linear optical network~\cite{tan2019resurgence,carolan2015universal} whose measurement output is used to implement logical gates by means of postselection and feedforward. Removing the feedforward, it is essentially equivalent to a quantum non-universal device known as the boson sampler~\cite{aaronson2011computational}. The boson sampler is an $M$-mode linear interferometer (realizing a unitary transformation $U(M)$) whose input modes are populated by single photons and the output is measured in the Fock basis. It rose to prominence after it was shown that under certain operating assumptions the boson sampler can claim quantum supremacy by sampling from the output probability distribution~\cite{aaronson2011computational} -- a task some believe to be beyond the capabilities of the fastest classical computers. Since then, much has been written on the topic~\cite{carolan2015universal,neville2017classical,clifford2018classical,brod2020classical,gg,zhong2020quantum}.

An interesting problem is whether the boson sampler can be turn into a near-term quantum device and make the alleged quantum supremacy useful. As a standalone device producing just measurement samples it seems unlikely but perhaps as part of a quantum-classical hybrid system equipped with an active feedback and a classical optimizer evaluating an objective function the chances are higher~\cite{chabaud2021quantum,shi2021quantum}. This type of general near-term device going under the name~\emph{variational quantum eigensolver} without active error-correction has been proposed for almost all non-photonic quantum computing platforms~\cite{campos1989quantum,bharti2021noisy}. The goal is to generate an expressive ansatz state with the help of entangling gates. The issue with photons is that they are bosons and not qubits. A priori it is not a problem. For example, we can use the dual-rail encoding to make them more qubit-like  but this somehow misses the point for the near-term variational circuits. Linear interferometers are capable of producing  highly entangled multipartite states but they cannot deterministically generate any (entangled) state we wish. Or, put differently, we cannot encode any qubit Hamiltonian like for generic variational circuits~\cite{campos1989quantum,bharti2021noisy}. The mechanism behind creating entangled states in linear bosonic circuits is quite different from the qubit circuits and it is directly related to the boson statistics as demonstrated by the Hong-Ou-Mandel experiment~\cite{hong1987measurement}.

Other issues faced by linear photonic circuits are more familiar from other platforms and among them the most prominent one is the occurrence of errors. The major source of errors in photonic circuits is photon loss. The depth of a linear optical circuit implementing $U(M)$ linearly increases with the mode number~$M$, causing the photon loss to increase exponentially. Passive error mitigation techniques, typically based on classical postprocessing, only postpone the inevitable loss of the quantum character of an output distribution, which makes the circuit efficiently classically simulable. Lacking error correction, probably the only scalable strategy to avoid the fate of drowning in noise is to keep the interferometer shallow.  Unlike full-depth random photonic circuits, there is much less known about them. Perhaps the most detailed study is~\cite{lubasch2018tensor}, where it is argued that a simulation of shallow, but not too shallow, bosonic circuits indeed remains classically intractable. A shallow optical circuit with a different geometry and the complexity of its classical simulation has also been studied, see~\cite{brod2015complexity}.

In this study we offer a solution to some of the issues accompanying near-term bosonic computers. First, being aware of the limitations of bosonic entanglement generated in linear circuits and the subsequent  Fock measurement, we come up with a method of mapping the measurement in the boson Hilbert space as if it was done in a many-qubit Hilbert space using the parity function. We demonstrate a number of its desirable properties and a very rich behavior despite its conceptual simplicity. We use it to solve instances of the Ising model, which is a purely classical model of particle interaction in statistical physics. In the optimization/operation research circuits the model is known as quadratic unconstrained binary optimization (QUBO) problem. The importance of QUBO from practical perspective cannot be overestimated~\cite{lucas2014ising,kochenberger2014unconstrained} but our method can be used even for other types of optimization problems, where a QUBO formulation may be cumbersome. One of them is the portfolio optimization problem with binary investment, which is known to be intractable~\cite{venturelli2019reverse}.
\begin{figure}[t]
  \resizebox{7cm}{!}{\includegraphics{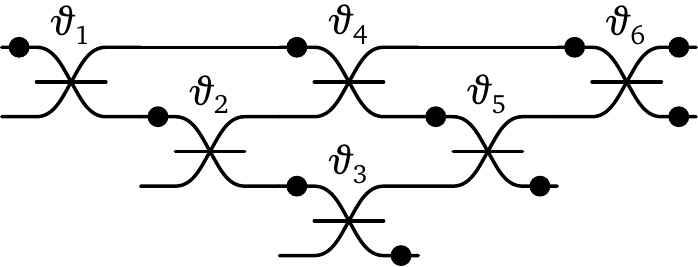}}
  \caption{Decomposition of $U(4)$ according to~\cite{reck1994experimental} into six beam-splitters parametrized by $\vt_i$ and ten phase-shifters represented by the black dots. }
  \label{fig:reck}
\end{figure}

To address the lack of error correction and fault tolerance we focus on a class of shallow bosonic circuits and the methods we use are borrowed, perhaps surprisingly, from lattice theory and enumerative combinatorics~\cite{stanley1997enumerative}. We identify certain partially ordered sets (posets) as the central object in uncovering the structure of bosonic circuits. Among them are finite Young's lattices and Boolean lattices. The developed methods are universal for any circuit geometry but our main focus is on a large family of shallow circuits derived from the triangular scheme by Reck et. al.~\cite{reck1994experimental}. In particular, we consider a sequence of $M$-mode shallow photonic circuits whose depth can be gradually increased all the way to a full-depth $M$-mode circuit whose example for $M=4$ is in Fig.~\ref{fig:reck}. We take a product of the elementary unitaries corresponding to circuits of an ever increasing depth. For the illustrated case, the first `slice' (the shallowest 1D circuit which is known to be classically efficiently simulated~\cite{lubasch2018tensor}) would be described by $\prod_{i=1}^{3}U(\vt_{4-i})$, the second, less shallow, circuit is $\prod_{i=1}^{5}U(\vt_{6-i})$ and the last one is the full circuit (we ignore the phases in this example).

The input product state of $n\leq M$ photons is limited from practical reasons to at most one photon per input mode. Besides the desirable theoretical properties of the $M$-mode Reck scheme described in detail in this paper, its practical advantage is a natural implementation as a time-bin interferometer with $M$ loops in series. This is a very tempting shortcut towards large-scale shallow bosonic circuits. But as we mentioned, our analysis is oblivious to any specific circuit implementation or depth and can be applied to the rectangular scheme~\cite{clements2016optimal} with minor, mostly technical, modifications as well.

The structure of the paper is the following. Section~\ref{s:generalSetup} is the bird's eye view of the studied variational bosonic circuit. In Section~\ref{s:basisMapping} we explain the main idea behind the proposed mapping and some of its basic properties. In Section~\ref{s:algo} we describe the variational bosonic algorithm in detail and present some supporting results such as the parameter shift rule. Section~\ref{s:apps} is dedicated to solved problems (QUBO, classical Ising problem and binary portfolio optimization) followed by brief discussions in Section~\ref{s:deeperCircuit} and Section~\ref{s:finitesampling} about the advantages of deeper circuits and the positive effects of finite sampling. Section~\ref{s:TBI} contains a description of how to implement the variational solver as a time-bin interferometer and in Section~\ref{s:challenges} we mention some challenges we face related to scalability. Section~\ref{s:techdetails} contains a detailed theoretical analysis. In Section~\ref{s:parityFcn} we introduce the parity function and the Hilbert spaces mapping in detail. Section~\ref{s:latticePath} introduces integer lattices and Dyck paths followed by Section~\ref{s:ferrers}, where we mention a few facts about lattice theory with an emphasis on two important objects known as Young's lattice and Boolean lattice. All comes together in Section~\ref{s:shallowCat}, where we link these concepts with shallow bosonic circuits and prove several theoretical results underpinning the studied applications. The most important among them is that  a measurement even in the case of the shallowest bosonic circuit can be mapped to a measurement in a qubit Hilbert space.

This paper is complemented by a public Github repository, where the presented examples (and much more) are shown~\footnote{\href{https://github.com/orcacomputing/quantumqubo/}{https://github.com/orcacomputing/quantumqubo}}.

\section{Solving interesting problems with a (shallow) boson sampler }\label{s:bsForApps}

\subsection{The general setup}\label{s:generalSetup}
The setup is not unlike other near-term quantum devices using classical variational algorithms, which aim at finding the minimum of a cost or objective function $f: \{0,1\}^{M} \rightarrow\bbR$~\cite{Moll_2018, liu2021layer}. An output from a quantum circuit is measured and the resulting data is fetched into a classical optimizer which evaluates the objective function. The circuit parameters are updated based on the evaluation and the cycle repeats until a satisfactory solution is found.

In the bosonic case, however, we face a serious obstacle. The interferometer multipartite output Fock state can't be called a multiqubit state in any  reasonable sense. The same holds for the Fock measurement -- it is quite dissimilar to a measurement in the qubit basis. Recall that a true qubit-based near-term variational circuit is typically measured in two complementary bases depending on what observable is investigated. Despite this handicap we developed a method, where the Fock states' measurement can be interpreted as a measurement in a~\emph{fixed} qubit basis as if happening in an abstract multi-qubit Hilbert space. This is a considerable advance in the applicability of a boson sampler as a near-term quantum device but we have to keep in mind that the fixed basis does not allow us to simulate just any Hamiltonian.

\begin{figure}[t]
	\resizebox{14cm}{!}{\includegraphics{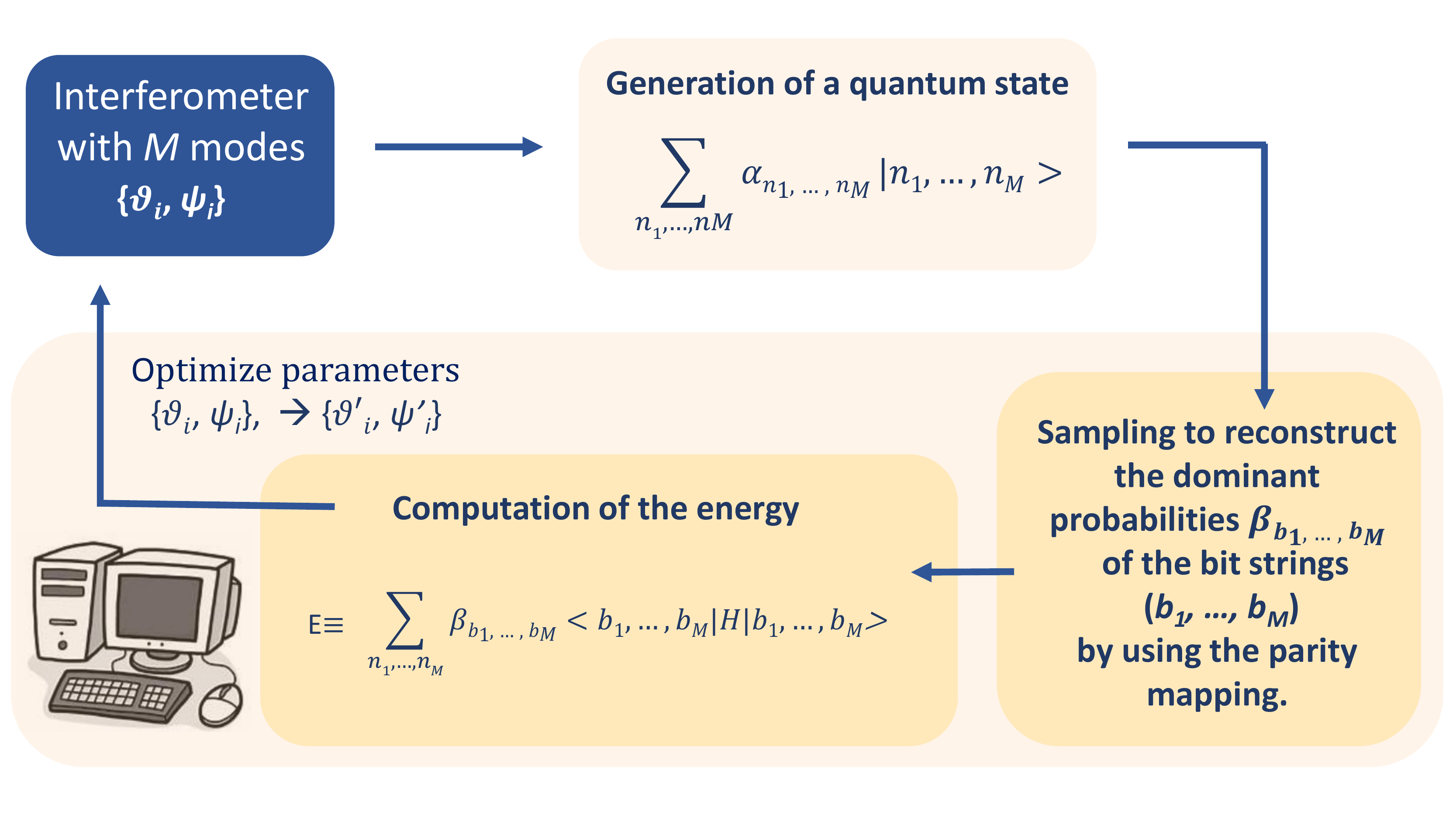}}
	\caption{A quick glance over the investigated variational bosonic solver. A quantum state is generated by an $M$-mode optical
    interferometer whose amplitudes are function of the beam splitter and phase shifter angles $\vt_{i}$ and $\psi_i$. The output state is measured in the Fock basis whose output is a detection pattern -- a string of non-negative integers $\n=(n_{1},\dots, n_{M})$. The detection pattern is mapped to a binary $M$-tuple $\bb^{(j)}=(b^{(j)}_{1},\dots,b^{(j)}_{M})$ by virtue of a parity function~$\wp_j$. Alternatively, the parity of each mode is measured directly. We repeat the measurement $N_s$ times to reconstruct the probability of the most likely states. We compute an objective function that corresponds to the sum of the energies of each bit string weighted by their probability (cf. Eq.~\eqref{eq:energy}). The objective function is then minimized by a classical optimizer providing an update to the interferometer parameters $\vt_{i},\psi_i$ and the cycle repeats until convergence is reached.}
	\label{fig:bosonicVQE}
\end{figure}

The overall bosonic setup is depicted in Fig.~\ref{fig:bosonicVQE}. The output modes of the interferometer are equipped with photon number resolving detectors (PNRs) and each measurement result is recorded as an $M$-tuple $\n=(n_1,\dots,n_M)$. For the initial run, the phase and beam-splitter angles are chosen at random or by a judicious choice. The circuit is sampled several times and the measurement results are recorded. After the quantum run comes the crucial classical postprocessing step whose details are explained in the next section.

\subsection{Qubit basis mapping}\label{s:basisMapping}

In Sec.~\ref{s:shallowCat} we introduce a sequence of the so-called Catalan Hilbert spaces $[\rC_i(M,n)]$, where each $\rC_i(M,n)$ is a subspace of a completely symmetric Hilbert space $\rH^+(M,n)$ of $n=\sum_{i=1}^Mn_i$ bosons in $M$ modes of dimension $|\rH^+(M,n)|=\binom{n+M-1}{n}$. The reason behind the adjective Catalan is the dimensionality of $\rC_i(M,n)$ related to the Catalan numbers. To intuitively introduce the Catalan Hilbert spaces, if the space $\rH^+(M,n)$ is addressed by the full-depth $M$-mode photonic interferometer then the Catalan space $\rC_i(M,n)$  is simply the Hilbert subspace of $\rH^+(M,n)$ accessed by the optical states generated by the first $i$ layers of the interferometer. Introduced just like that, the concept of a Catalan Hilbert space is still too general. By $\rC_i(M,n)$ we specifically refer to the first $i$~layers of the triangular Reck scheme~\cite{reck1994experimental}. There is still some freedom left in the way the  $n$~input photons can be distributed. Motivated by practical considerations we focus on $n\leq M$ and, furthermore, we focus on the regime of at most one photon per mode. There is one more constraint we describe later in this section.

 We introduce a function (in fact two closely related functions) which maps a set of basis states of $\rC_i(M,n)$ or $\rH^+(M,n)$ to a set of basis state of an abstract many-qubit Hilbert space and investigate its properties in detail. Let $\wp_j:(n_1,\dots,n_M)\mapsto (b_1^{(j)},\dots,b^{(j)}_{M'})$ defined as
\begin{equation}\label{eq:parityFcn}
  \wp_j: b_i^{(j)}=\bmod{[n_i,2]}\oplus j,
\end{equation}
where $b_i^{(j)}\in\{0,1\}$,  $j=0,1$ and $\oplus$ denotes binary addition. The qubit number $M'$ is closely related to~$M$ as we will see shortly. We can understand $\wp_0$ as a component-wise parity function and $\wp_1$ as its additive inverse. Let's make an identification $\n=(n_1,\dots,n_M)\leftrightarrow\ket{n_1,\dots,n_M}$, where the latter denotes a bosonic Fock basis state. The right number of such states spans $\rC_i(M,n)$ or $\rH^+(M,n)$. We identify the bit string $\bb^{(j)}=(b_1^{(j)},\dots,b_{M'}^{(j)})$ obtained from~\eqref{eq:parityFcn} with the standard basis $\ket{b_1^{(j)}\dots b_{M'}^{(j)}}$ of an $M'$-qubit Hilbert space $\H_{M'}$ of dimension $2^{M'}$. It is not a priori clear how $\wp_j$ `performs'. The Hilbert space dimensions are incompatible no matter what $M'$ is. Our main result in Sec.~\ref{s:parityFcn} is to show the relationship between $M$ and $M'$ and that $\wp_j$ is surjective, namely the non-obvious part, where none of the standard basis states of $\H_{M'}$ is left out (naturally, all bases of  $\rH^+(M,n)$ map to something in $\H_{M'}$ by virtue of $\wp_j$ so it is a well-defined function). Similarly for all $\rC_i(M,n)$, see~Sec.~\ref{s:shallowCat}.

Perhaps we can start with what $\wp_j$ is \emph{not}.  Let $\n,\n'$ be two measurement patterns. If both are mapped to the same bit string~$\bb^{(j)}$ then their probabilities are summed. So what is happening is that if the bosonic state is $\a\ket{n}+\b\ket{n'}$ then the probability of measurement of $\ket{n}$ or $\ket{n'}$ is $|\a|^2+|\b|^2\df~c$. Since $\wp_j$ maps both states to $\ket{b^{(j)}}$ it is as if its amplitude was $\sqrt{c}$ (note that we are loosing the phase information, or more precisely, we can choose any phase we wish). So the functions $\wp_j$ are not a linear map from $\rH^+(M,n)$ (or $\rC_i(M,n)$) to $\H_{M'}$ let alone an isometry: $\wp_j(\a\ket{n}+\b\ket{n'})=(\a+\b)\ket{b^{(j)}}\neq\sqrt{c}\ket{b^{(j)}}$). But this is fine for our purposes since we do not wish to coherently map one Hilbert space to another, trying to preserve everything there is -- we are just grouping together certain measurement results and interpret them as a measurement in~$\H_{M'}$.

Our other main result of Sec.~\ref{s:parityFcn} is to show that $M'=M$ even for the shallowest circuit $\rC_i(M,n)$. But for this to be true we have to sample the circuit for $n=M$ and $n=M-1$ 
and in both cases apply $\wp_j$ for $j=0,1$. Hence there are in total four sampling steps for any $M$ and for circuits of any depth. There is an option to reduce the number of sampling steps to two or even one at the expense of mapping to the $(M-1)$-qubit basis state set (hence $M'=M-1$). The details depend on the parity of $M$ as we discuss in Sec.~\ref{s:parityFcn}. But our  analysis suggests that such options provide slightly worse practical results and will not be considered here.

Before we go on we offer three comments. There exists an alternative quantum setup to what we just described. Instead of the PNRs followed by~\eqref{eq:parityFcn} we could use the so-called parity measurement~\cite{plick2010parity,gerry2010parity} for each mode which merges the detection step with the parity function. Even though the performance of the PNRs has steadily improved over the past decades they are still relatively costly pieces of an equipment and may be slow, bulky, noisy, limited in the number of photons that they can resolve and any combination thereof. The parity measurement, on the other hand, is based on a homodyne measurement and this is a very mature technology.

What is the overall motivation behind using the parity function? We would like to pretend that the Fock measurement in the bosonic Hilbert space is as if we are sampling a system of many abstract entangled qubits. As noted before, there are limits to this given by the fact that the Fock measurement is a fixed-basis measurement. On the other hand, there are interesting computational tasks where it does not matter and some of them are presented in Sec.~\ref{s:apps}. We have found a way  to mimic a measurement in a qubit system by virtue of Eq.~\eqref{eq:parityFcn} but this is not sufficient per se. An equally important purpose of the parity function is to coarse-grain the classical data obtained by sampling a bosonic device. When comparing the dimension of $\rH^+(M,n)$ for $n=M,M-1$ with the $M$-qubit Hilbert space we indeed see a huge (exponential) redundancy on the bosonic side. In other words, many bosonic bases can be mapped to a single $M$-qubit basis and indeed the parity function does it in a desirable way, see Sec.~\ref{s:parityFcn}. A similar result holds for~$\rC_i(M,n)$ even though it is not solved as satisfactorily as for the full depth circuit, see Sec.~\ref{s:shallowCat}.
\begin{figure}[t]
    \resizebox{15cm}{!}{\includegraphics{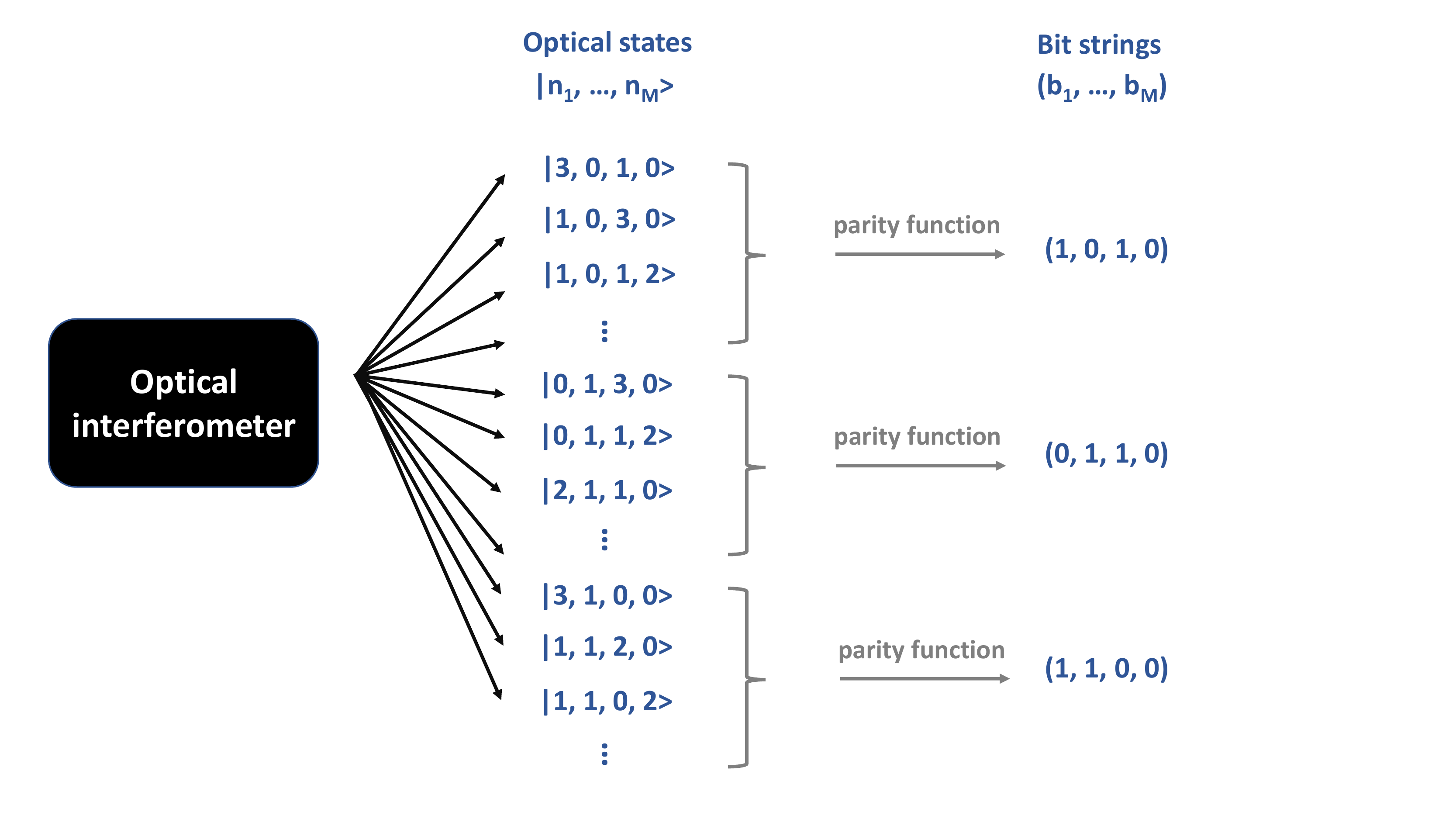}}
	\caption{Visualisation of parity mapping $\wp_0$ for $n=M$. The optical circuit generates states spanned by the basis of the $i$-th ``shallow
    Hilbert space'' $\rC_i(M,n)$. We then classically map those states to bit strings $\bb^{(0)}=(b^{(0)}_1,\dots, b^{(0)}_M)$ using the parity function $\wp_0$. A noteworthy property of this configuration is that the Hilbert space spanned by the bosonic basis grows exponentially faster than the one spanned by the bit strings (whose dimension is obviously exponential as well). Therefore, multiple photonic detection events will lead to a single bit string and we show that no bit string is omitted for any $\rC_i(M,n)$, where $n=M,M-1$.}
	\label{fig:MappingPle}
\end{figure}

Our last comment is to emphasize that the parity function is unlike the action of the so-called threshold detector. The threshold detectors also output zeros and ones according to whether  zero photons are measured or not. The methods developed in Sec.~\ref{s:techdetails} can be used to show whether the `threshold map' has similar properties to the parity map even for the shallowest circuits.

\subsection{Variational bosonic solver}\label{s:algo}

In this section we describe the variational bosonic solver whose overall scheme is in Fig.~\ref{fig:bosonicVQE}. The central part is the construction of the objective function we want to minimize from the estimated probabilities of photon detection patterns $\n=(n_1,\dots,n_{M})$, where $n_i\geq0$ is the number of photons in mode~$i$. The novelty here is the ability to map certain computational problems typically formulated in an abstract qubit Hilbert space to the measurement in a bosonic Hilbert space.
\begin{algorithm}
	\caption{Variational bosonic solver}
	\label{algo:1}
	\begin{algorithmic}
		\STATE \textbf{Input}:
		\STATE Number of modes $M$.
		\STATE Number of samples $N_s$.
		\STATE Initial parameter $E_{ini}$ that is much bigger than any other value of the problem.
		\STATE \textbf{Output}:
		\STATE Minimal energy $E_{min}$.
		\STATE Configuration of minimal energy $\bb_{min}$.
		\STATE
		\STATE $E_{min} \gets E_{ini} $
		\FOR{$n=M$ and $n=M-1$}
		\FOR{parity $\wp_j$, $j \in \{0,1\}$}
		\STATE $E(j;\bvt,\bpsi) \gets 0$
		\STATE $\vt,\psi$ are initialised with random values.
		\FOR{$k \gets 1$ to $N_s$}
		\STATE We measure the output pattern $\n=(n_1, \dots, n_{M})$ generated by the interferometer.
		\STATE We classically map $\n$ to bit strings $\bb^{(j)}=(b^{(j)}_1, \dots, b^{(j)}_{M})$ by using the parity function $\wp_j$:
		\STATE $\bb^{(j)} = \wp_j(\n)$
		\STATE Compute the energy of the bit string: $E_{\bb}^{(j)} = \bra{b^{(j)}_{1}, \dots, b^{(j)}_{M}}H\ket{b^{(j)}_1,\dots,b^{(j)}_{M}} $
		\STATE $E(j;\bvt,\bpsi) \gets E(j;\bvt,\bpsi) + E_{\bb}^{(j)}/{N_s}$
		\IF{$E^{(j)}_{\bb} < E_{min}$}
		\STATE $E_{min} = E^{(j)}_{\bb}$
		\STATE $\bb_{min} = \bb^{(j)}$
		\ENDIF
		\ENDFOR
		\STATE Update $\vt,\psi$ to minimize $E(j;\bvt,\bpsi)$.
		\STATE Iterate until convergence is reached.
		\ENDFOR
		\ENDFOR
	\end{algorithmic}
\end{algorithm}
The state generated by the interferometer can be written as
\begin{equation}\label{eq:state}
	\ket{\xi(\bvt,\bpsi)} = \sum_{\n} \alpha_{\n} \ket{n_1, \dots, n_M},
\end{equation}
where  $\bvt$ is a collection of beam-splitter angles~$\vt_j$ and $\bpsi$ a collection of phases $\psi_j$. Their total number depends on the depth of the circuit. We sample from the interferometer $N_s$ times and map the detection patterns to bit strings using the parity mapping $\wp_j$, $j \in \{0,1\}$ described by Eq.~\eqref{eq:parityFcn} and coarse-grain the probabilities:
\begin{equation}\label{eq:f}
	\text{parity function}\ \wp_j\ \ \left\{
	\begin{aligned}
		\n=(n_1,\dots, n_M)  &\rightarrow \bb^{(j)}=(b^{(j)}_1,\dots, b^{(j)}_{M}) \\
		 p_{\n} &\rightarrow \b_{\bb^{(j)}}\in[0,1]
	\end{aligned}
	\right.
\end{equation}
Each bit string $\bb^{(j)}$ is then accompanied by its estimated probability $\b_{\bb^{(j)}}$. For example, the measured states $\ket{3,0,1,0}$ and $\ket{1,2,1,0}$ both lead to the bit string $\bb^{(0)}=(1,0,1,0)$ for $\wp_0$ and  by counting the number of occurrences of the bit string the probability is estimated, see Fig.~\ref{fig:MappingPle}. Our objective function for an observable $H$ is then
\begin{equation}\label{eq:energy}
E(j;\bvt,\bpsi)\df\sum_{\bb^{(j)}} \b_{\bb^{(j)}} \bra{b^{(j)}_1, \dots, b^{(j)}_M}H\ket{b^{(j)}_1, \dots, b^{(j)}_M}.
\end{equation}
As we iterate, we keep track of the configuration that led to the smallest energy. In theory, such configuration should appear in the final distribution. However, in practice finite sampling leads to noise that can prevent the algorithm from sticking to a new low energy configuration. The number of bit string energy evaluations made by Algorithm~\ref{algo:1} is $8 \times N_{\text{parameters}} \times N_{\text{iterate}} \times N_s$, where the factor 8 comes from the fact that each gradient requires 2 evaluations of Eq.~\eqref{eq:energy} and that we do four gradient descents in total. The algorithm is formalized on p.~\pageref{algo:1}.

\subsubsection*{Parameter shift rule derivation}
At each iteration, one also needs to compute the gradient of $E$ w.r.t. each variable while keeping the rest constant in order to apply the usual gradient descent rule:
\begin{equation}\label{eq:descent}
	\vt'_{i} \rightarrow \vt_{i} - \eta \frac{\partial E}{\partial \vt_{i}},
\end{equation}
where $\eta$ is the learning rate. Updating the angles $\vt_{i}$ changes the probability of the state generated by the optical circuit, which will have an impact on the values of the $\b_{\bb}$. Computing~\eqref{eq:descent} by using the numerical approximation
\begin{equation}\label{eq:gradientnumerical}
\frac{\partial E}{\partial \vt_{i}} \simeq \frac{E(\vt_{i} + \epsilon) - E(\vt_{i}) }{\epsilon}
\end{equation}
is experimentally questionable  since it would require not only an $\epsilon$-accuracy in the way we tune the optical components but also enough samples to capture the differences between the very similar distributions $E(\vt_{i}+ \epsilon)$ and $E(\vt_{i})$. Instead, a trick called the \emph{parameter shift rule}
\begin{equation}\label{eq:gradAnal}
  2\frac{\partial E(\vt_i)}{\partial \vt_i}=E(\vt_i+\pi/2)-E(\vt_i-\pi/2),
\end{equation}
first reported in~\cite{li2017hybrid,mitarai2018quantum} can be used. To show the validity of the rule for~\eqref{eq:energy}, consider the following function
\begin{equation}\label{eq:Qsampled}
  f(\bvt,\bpsi)=\bra{in}U^\dg(\bvt,\bpsi)OU(\bvt,\bpsi)\ket{in},
\end{equation}
where $\ket{in}$ is the input state of the interferometer, $O$ is an observable and $U(\bvt,\bpsi)=\prod_{k}U_{ij}(\vt_k,\psi_k)$ is a product of the building blocks coupling the modes $i$ and $j$ (beam-splitters and phase shifters like in the Reck mesh depicted in Fig.~\ref{fig:reck}), where
\begin{equation}\label{eq:BSandPS}
  U_{ij}(\vt_k,\psi_k)=\exp{[{\vt_k\over2}(a_i a_{j}^\dg-a_i^\dg a_{j})]}\exp{[-{\psi_k\over2}(a_i^\dg a_{i}-a_j^\dg a_{j})]}.
\end{equation}
Using the identity~\cite{campos1989quantum}
\begin{equation}\label{eq:SchrdHeis}
   U_{ij}(\vt_k,\psi_k)\begin{bmatrix}
                  a_i \\
                  a_j \\
                \end{bmatrix}
    U_{ij}^\dg(\vt_k,\psi_k)
    =e^{-i\psi_k/2}\begin{bmatrix}
       \cos{\vt_k\over2}e^{i\psi_k} & \sin{\vt_k\over2}e^{i\psi_k} \\
       -\sin{\vt_k\over2} & \cos{\vt_k\over2} \\
     \end{bmatrix}
     \begin{bmatrix}
                  a_i \\
                  a_j \\
     \end{bmatrix}
\end{equation}
we deduce
\begin{equation}\label{eq:SchwingerTransform}
  U_{ij}(\vt_k,\psi_k)(a_i^\dg a_j)U_{ij}^\dg(\vt_k,\psi_k) =  \big(\cos{\vt_k\over2}e^{-i\psi_k}a_i^\dg+\sin{\vt_k\over2}e^{-i\psi_k}a_j^\dg\big)\big(-\sin{\vt_k\over2}a_i+\cos{\vt_k\over2}a_j\big).
\end{equation}
With the help of the double-angle trigonometric formulas ($\cos^2{\vt/2}=(1+\cos{\vt})/2,\sin^2{\vt/2}=(1-\cos{\vt})/2$ and $2\cos{\vt/2}\sin{\vt/2}=\sin{\vt}$) we rewrite the RHS of~\eqref{eq:SchwingerTransform} as a sum of elementary trigonometric functions and so the parameter shift formula holds. Similarly, it holds for $\psi$ by using $e^{-i\psi}=\cos{\psi}-i\sin{\psi}$. A nearly equivalent result is valid for $U_{ij}(\vt_k,\psi_k)(a_i^\dg a_i)U_{ij}^\dg(\vt_k,\psi_k)$ and therefore for any observable of the form
\begin{equation}\label{eq:SchwingerObs}
  O=\sum_{i=1}^{M}o_{ii}a_i^\dg a_i+\sum_{i,j=1}^{M}o_{ij}a_i^\dg a_j.
\end{equation}
Expression~\eqref{eq:energy} becomes~\eqref{eq:Qsampled} after we rewrite the observable~$H$ of~\eqref{eq:energy} in~the Schwinger bosonic representation:
\begin{equation}\label{eq:Qmap}
H\mapsto[a_1^\dg,\dots,a_M^\dg]^\top H[a_1,\dots,a_M].
\end{equation}
Hence the parameter shift rule holds for~\eqref{eq:energy}. Note that unlike~\cite{schuld2019evaluating} we don't rely on any special matrix algebra properties or Gaussian evolution.

\clearpage

\subsection{Applications}\label{s:apps}
\subsubsection*{QUBO}
\begin{figure}[b]
	\resizebox{14cm}{!}{\includegraphics{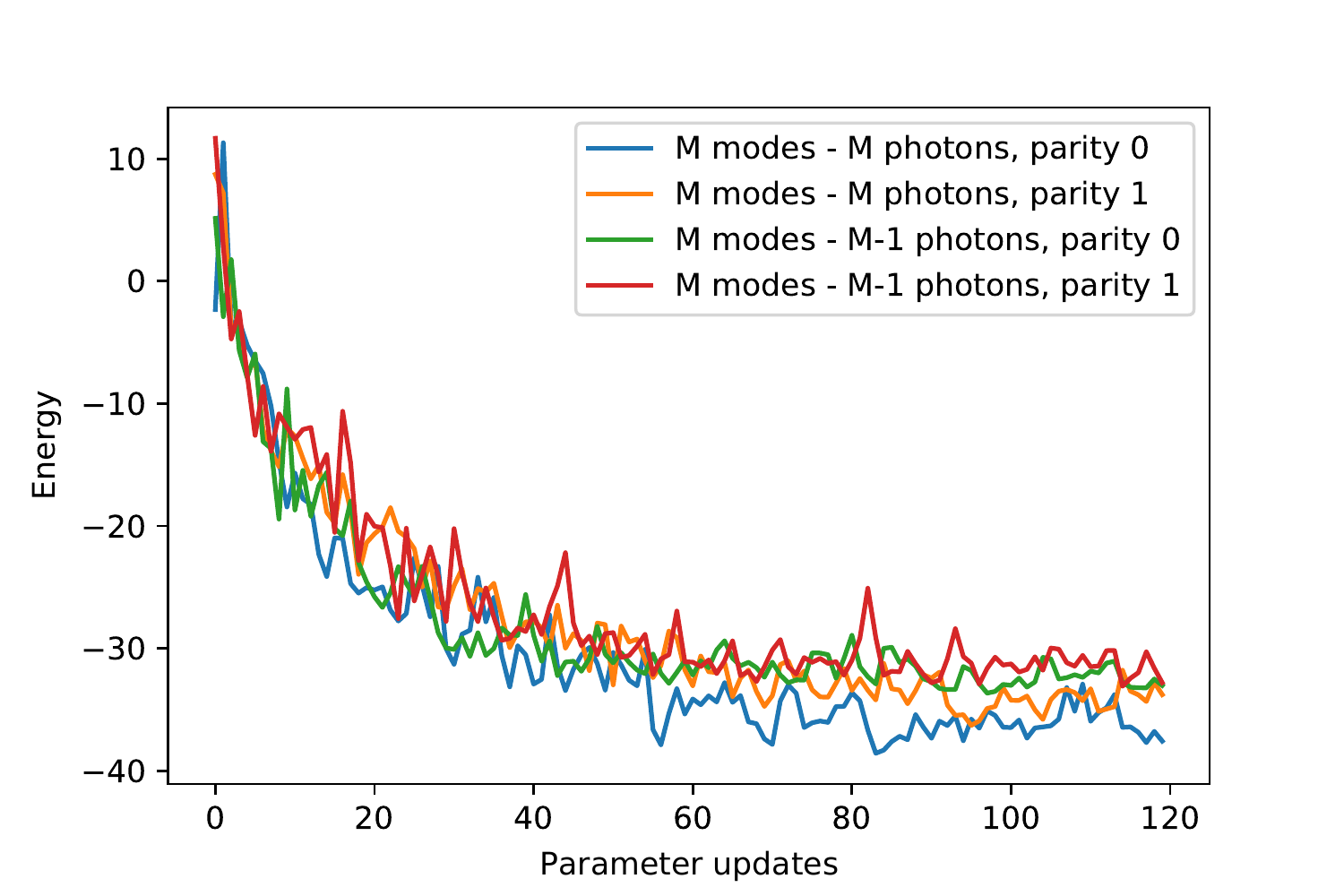}}
	\caption{Example of a QUBO problem with a $30 \times 30$ random symmetric matrix $Q$ in Eq.~\eqref{eq:QUBOopti}. We plot the energy of the system (Eq.~\eqref{eq:energy}) as a function of the number of iterations. The minimum found by the algorithm has an energy of $E_{min}=-41.43$, which corresponds to the theoretical minimum found by brute force computation ($2^{30} \simeq 10^9$ configurations).} 
	\label{fig:randomQUBO2}
\end{figure}

In the following, we focus on the QUBO optimization problem. QUBO is usually formulated~\cite{kochenberger2014unconstrained} as the following quadratic program:
\begin{equation}\label{eq:QUBOopti}
	\min_{\bs x}{[{\bs{x}^\top Q\bs{x}}]},
\end{equation}
where $Q\in\bbR^{M\times M}$ can be written as a symmetric matrix and $\bs{x}$ is an $M$-tuple such that $x_i\in(0,1)$. Finding an optimal solution to a QUBO problem is equivalent to minimizing a classical Ising Hamiltonian~\cite{lucas2014ising}
\begin{equation}\label{eq:IsingHamforQ}
	H=\sum_{1\leq i<j\leq n}J_{ij}s_is_j+\sum_{k=1}^nh_ks_k+const
\end{equation}
via the linear transformation $s_i=2x_i-1$, where $J_{ij},h_k\in\bbR$. We write $H=Q$ as a bilinear expression~\eqref{eq:Qmap} in terms of the field operators and minimize~\eqref{eq:energy}. In Fig.~\ref{fig:randomQUBO2} we show an example of a QUBO problem for a random symmetric matrix~$Q$ of dimension $M=30$.

\subsubsection*{Ising Hamiltonian on M\"obius graph}
\begin{figure}[t]
	\resizebox{14cm}{!}{\includegraphics{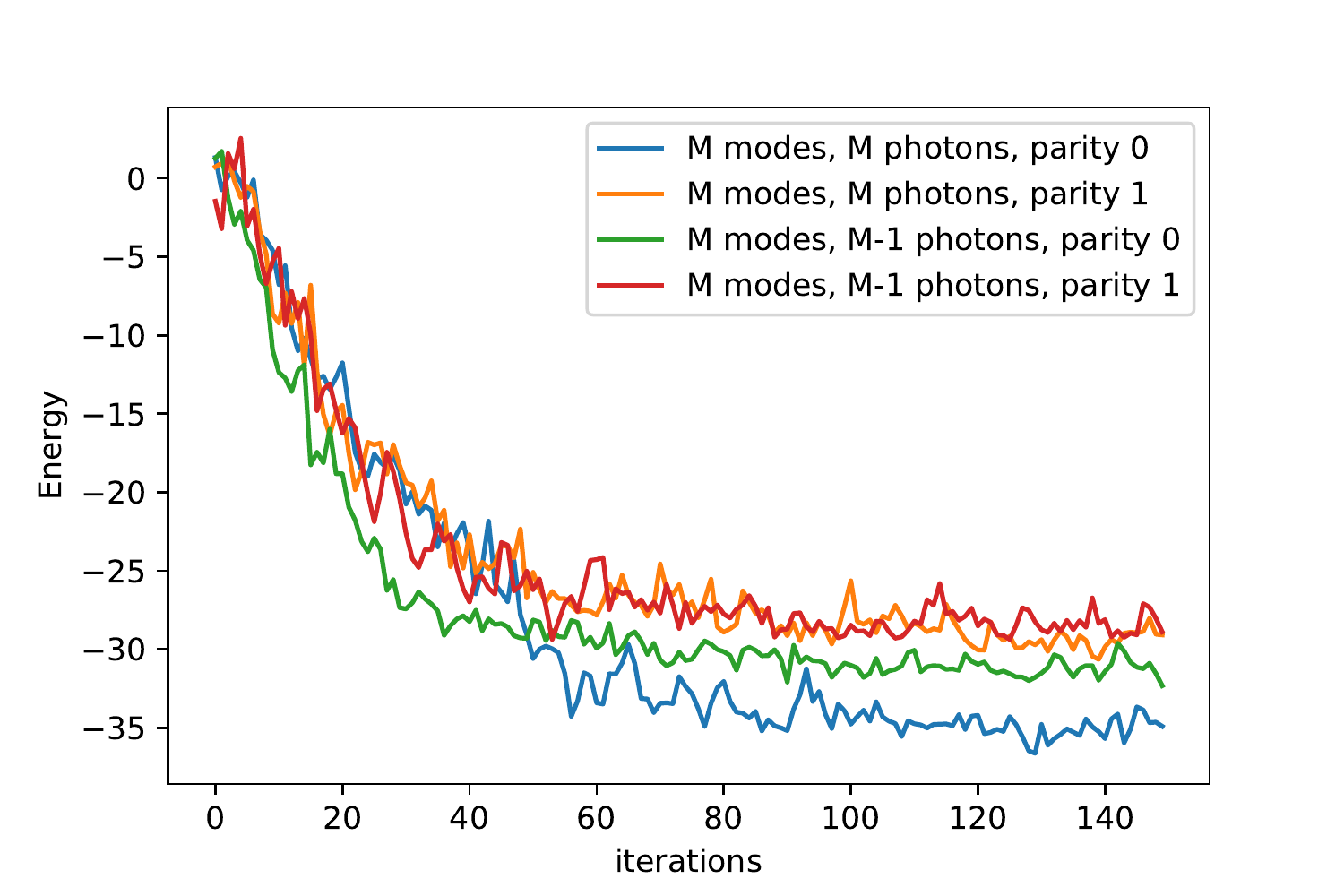}}
	\caption{Illustrating the variational bosonic solver for the M\"obius Hamiltonian in~\eqref{eq:Mobius} by sampling from $\rC_1(70,n)$ (the
			shallowest circuit with 70 modes and $n=69,70$ input photons). (Top) 4 learning curves corresponding to $n=69$ photons for both parity functions (green and red) and 70 photons for both parity functions (blue and orange) for the number of samples $ N_s = 150$. The lowest energy state found by the solver is $E=-39.6$, whereas the theoretical minimum of~\eqref{eq:solMobius} for $J_a = 0.5, J_b = -0.2$ is $E_{min}=-40$.}
	\label{fig:moebius}
\end{figure}
The following Hamiltonian family
\begin{equation}\label{eq:Mobius}
  H(n,J_a,J_b)=-J_a\sum_{i=0}^{n-1}s_is_{i+1}-J_b\sum_{i=0}^{n/2-1}s_is_{i+n/2},
\end{equation}
whose defining graph is a ``twisted'' ladder by imposing the periodic boundary condition $s_{n} = s_0$, has an analytical solution in the form
\begin{equation}\label{eq:solMobius}
  \min{H(n,J_a,J_b)}=\min{[-nJ_a-nJ_b/2,(4-n)J_a+nJ_b/2]}
\end{equation}
whenever $J_a\in\bbR^+,J_b\in\bbR$.
The top graph of Fig.~\ref{fig:moebius} shows the necessity of running the experiment in four configurations in order to span the whole qubit Hilbert space of dimension~$2^M$.  We see four learning curves as we iterate the gradient descent. Whereas the best solutions found by the red, orange and green curves is only $E=-34$, the configuration $n=M$ with parity 0 (the blue curve) manages to land on a solution  $E=-39.6$ very close to the theoretical minimum  $E_{min}=-40$ verified by brute-force calculated minimal energy of~\eqref{eq:solMobius}. As demonstrated in Sec.~\ref{s:deeperCircuit} we have a good reason to believe that by sampling a  deeper circuit we may further approach the global minimum solution.

\subsubsection*{Portfolio optimization}
\begin{algorithm}
	\caption{Variational bosonic solver -- portfolio optimization with binary investment}
	\label{alg:2}
	\begin{algorithmic}
		\STATE \textbf{Input}:
		\STATE Number of  companies $N$.
		\STATE Returns $\mu$ of those $N$ companies.
		\STATE Covariance $\Sigma$ of those $N$ companies.
		\STATE Paremeters $E_{ini},E_{pen}$ that are much bigger than any other value of the problem.
		\STATE \textbf{Output}:
		\STATE Minimal energy $E_{min}$.
		\STATE Configuration of minimal energy $\bb_{min}$.
		\STATE
		\STATE $E_{min} \gets E_{ini}$
		
		\FOR{$n=M$ and $n=M-1$}
		\FOR{parity $\wp_j$, $j \in \{0,1\}$}
		\STATE Sample $N_s$ detection patterns and map them to bit strings $\bb^{(j)}=(b^{(j)}_1, \dots, b^{(j)}_{M}$) according to~\eqref{eq:f}.
		
		\STATE $ E \gets 0$
		\FOR {each bit string sampled $\bb^{(j)}$}
		\IF{$\sum_ib_i^{(j)} = 0$}
		\STATE $E \gets E + E_{pen}$
		\ELSE
		\STATE Normalize $\bb^{(j)}$ : $\bb^{(j)} \gets \frac{\bb^{(j)}}{\sum_i b^{(j)}_i }$
		\STATE $E^{(j)}_{new} = {\bb^{(j)}}^T \mu - \gamma \bb^{(j)} \Sigma \bb^{(j)}$
		\STATE $E \gets E + E^{(j)}_{new}$
		\IF{$E^{(j)}_{new} < E_{min}$}
		\STATE $E_{min} \gets E^{(j)}_{new}$
		\STATE $\bb_{min} \gets \bb^{(j)}$
		\ENDIF
		\ENDIF
		
		\ENDFOR
		
		\STATE Update $\vt,\psi$ to minimize $E^{(j)}(\bvt,\bpsi)$.
		\STATE Iterate until convergence.
		\ENDFOR		
		\ENDFOR
		
	\end{algorithmic}
\end{algorithm}

Our  variational bosonic solver can deal with other types of optimization problems besides QUBO. A problem of a great importance in financial risk assessment is known as portfolio optimization. It consists of finding an investment spread among $N$ fixed assets such that it maximizes the returns for a given risk. The key idea is that the standard deviation of a portfolio made of $N$ assets, i.e., the risk, is not the sum of the $N$ standard deviations of each asset taken individually. This means that diversifying by taking into account correlations between the assets can decrease the portfolio risk. This idea was first theorised by Markovitz~\cite{Markowitz1952}.

In this section, we consider $N$ assets that have returns $\mu_i, 1 \leq i \leq N$. We denote $\Sigma$ to be the $N$-dimensional covariance matrix between the returns of these assets. If one invests a proportion $\omega_i$ of its total investment in the asset~$i$, then the return $\mu_p$ of the portfolio containing all those assets is
\begin{equation}\label{eq:portfolioReturn}
	\mu_p =  \sum_{i=1}^{N} \omega_{i} \mu_{i} = \omega^{\top} \mu
\end{equation}
and its risk
\begin{equation}\label{eq:portfolioRisk}
	\s_p^2 = \sum_{1 \leq i,j \leq N} \om_{i}\omega_{j} \Sigma_{ij} = \om^{\top} \Sigma \om,
\end{equation}
where $\omega = \{\omega_i\}_{1 \leq i \leq N}$ represents the proportion of the total investment in each asset. We thus have $\omega \in [0,1]$. In the case of a static portfolio, the aim is to minimize the following function:
\begin{equation}\label{eq:hamiltonianPortfolio}
	f(\om) = - \om^{\top} \mu + \gamma \om^{\top}\Sigma\om
\end{equation}
with the additional constraint $\sum_i\om_i=1$. The parameter $\gamma$ is the risk aversion of the investor. When $\omega$ is continuous, one can use Eq.~\eqref{eq:hamiltonianPortfolio} to find the optimal investment~\cite{mugel2020dynamic}. However, when $\omega$ is discrete this problem is known to be intractable.

Let's describe the data preparation procedure. The daily expected return of a company $i$, $\mu_i$, is obtained by computing the average variation of the stock between consecutive days
\begin{equation}\label{eq:returndata}
	\mu_i^{\text{day}} = \Bigg \langle \log\left[\frac{S_{i}(T)}{S_{i}(T-1)}\right] \Bigg \rangle _{T},
\end{equation}
where $S_{i}(t)$ is the stock of the company at time~$t$. The presence of a logarithm to compute the daily return is due to the approximation
\begin{equation}\label{eq:apprxlogreturn}
	\frac{S_{i}(T) - S_{i}(T-1)}{S_{i}(T-1)} \simeq \log\left[\frac{S_{i}(T)}{S_{i}(T-1)}\right],
\end{equation}
which is the case for small daily variations. In Eq.~\eqref{eq:returndata}, the average can be done over any period of time~$T$. In practice, to express returns on an annual basis, one computes $\mu_{i}^{\text{year}} = \mu_i^{\text{day}} \times 250$ (there are 250 opening days per year). The covariance of the companies are also computed from the daily returns. We then multiply this covariance by 250 in order to express the variances annually. Here we compute the returns and the covariance matrix classically by using one year of daily returns. To encode $\omega$ into bit strings, we can discretize $\omega$ and perform a binary-encoding operation~\cite{mugel2020dynamic}
\begin{equation}\label{eq:portfolioEncoding}
	\omega_{i} = \frac{1}{2^{N_{q}} - 1} \sum_{q=0}^{2^{N_{q}} - 1} q x_{iq}.
\end{equation}

As mentioned previously, we do not formulate portfolio optimization as a QUBO problem. In fact, we can illustrate a limitation that appears when adding a constrain into a QUBO problem. In the case of portfolio optimization the constraint is $\sum_i\om_i=1$.
\begin{description}
    \item[Approach 1 -- QUBO formulation] Quantum computing platforms that must formulate an optimization problem as
    a QUBO instance would transform Eq.~\eqref{eq:hamiltonianPortfolio} into:
	\begin{equation}\label{eq:hamiltonianPortfoliowithconstrain}
		f(\omega) = - \omega^{\top} \mu + \gamma \omega^{\top} \Sigma \omega + B \big(\sum_i\om_i-1\big)^2,
	\end{equation}
	where $B$ is a constant chosen to be much larger than any parameters of the problem. This additional terms penalizes the terms that do not
    require the condition $\sum_i\om_i=1$. This approach means that the solver will explore a space of dimension $2^N$ whereas the space of valid solutions is a much smaller subspace of bit strings that satisfy $\sum_i\om_i=1$. In the case of $N=20$ companies, whose weights are encoded into $N_q = 3$ bits, the solver will explore a space of dimension $2^{60} \simeq 10^{18}$ but the subspace of the weights configurations satisfying $\sum_i\om_i=1$ has only  dimension 3168. With $\gamma = 1$, solving Eq. ~\eqref{eq:hamiltonianPortfoliowithconstrain} as QUBO leads to a minimum of $E=-0.2109$, whose Sharpe ratio is equal to 4.51 (Eq.~\eqref{eq:hamiltonianPortfolio} with $\gamma = 1$ is equivalent to finding the maximal Sharpe ratio.). The learning curves of that approach are shown in Fig.~\ref{fig:approach} on the left. The first 30 iterations have very high energies which shows how much the solver struggles staying in the subspace of the bit strings satisfying the constraint.

    \item[Approach 2 -- Non-QUBO formulation] The approach presented here is more suitable than QUBO. We can
    optimize the following function:
	\begin{equation}\label{eq:hamiltonianPortfoliononlinear}
		g(\omega) = - \frac{\omega^{\top}}{\sum_i\om_i} \mu + \gamma \frac{\omega^{\top} \Sigma \omega}{(\sum_i\om_i)^2}, \quad \forall
        \omega \quad \text{such that} \quad \sum_i\om_i \neq 0.
	\end{equation}
	If $\sum_i\om_i=1$, we return a higher value  than any other value of the problem. The advantage of this approach is that we will train
    the algorithm after normalizing the candidate solutions. This means that the space of acceptable solutions is of dimension $2^N$, which is much bigger than in Approach~1. Running the same experiment as in Approach~1, but by minimizing Eq.~\eqref{eq:hamiltonianPortfoliononlinear}, leads to a minimum of $E_{min}=-0.2217$. This corresponds to the Sharpe ratio equal to $6.68$. The learning curves of that approach are shown in Fig.~\ref{fig:approach} on the right.
\end{description}
\begin{figure}
	\label{fig:approach}
	\begin{subfigure}[h]{0.49\linewidth}
		\label{fig:approach1}
		\includegraphics[width=\linewidth]{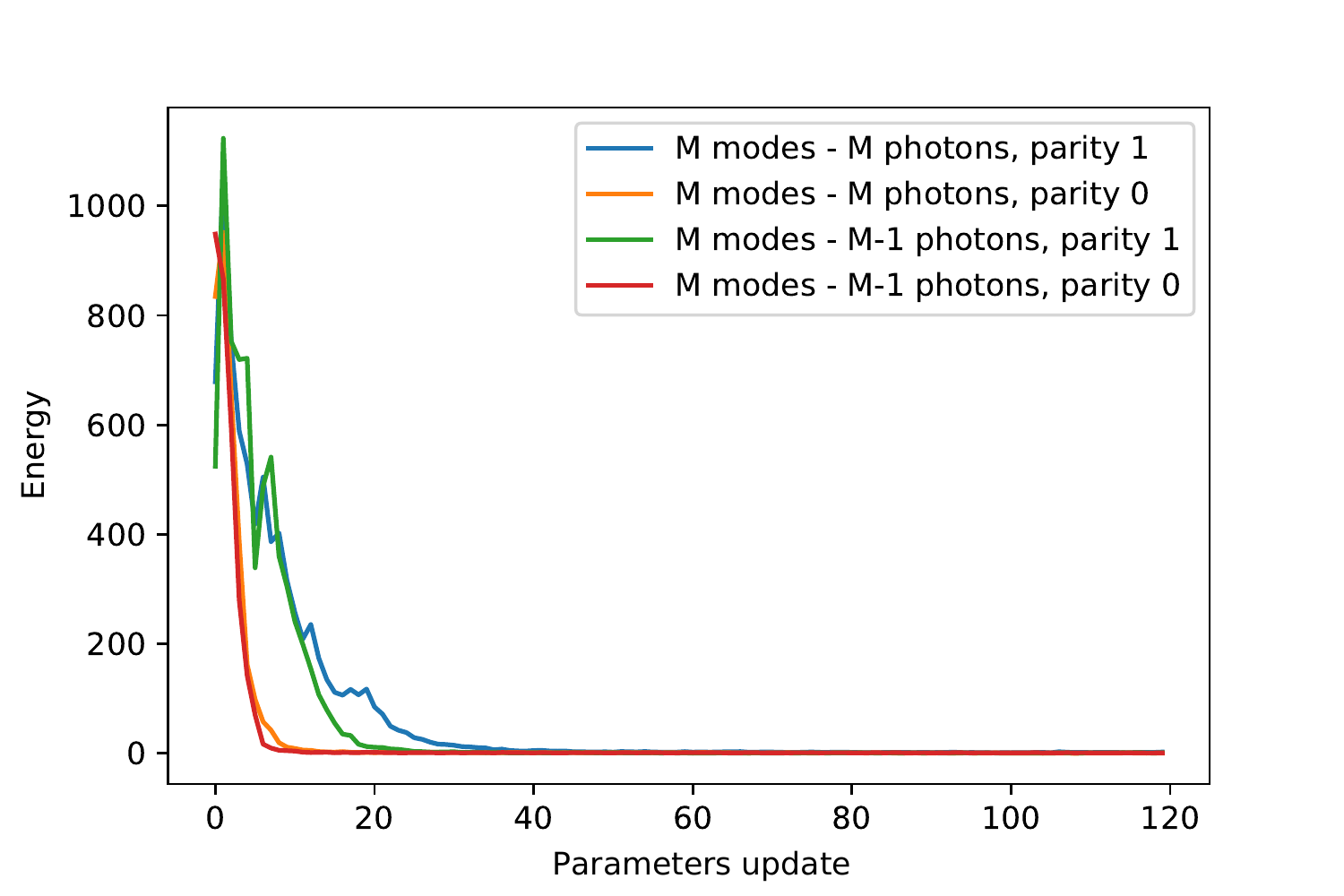}
	\end{subfigure}
	\hfill
	\begin{subfigure}[h]{0.49\linewidth}
		\label{fig:approach2}
		\includegraphics[width=\linewidth]{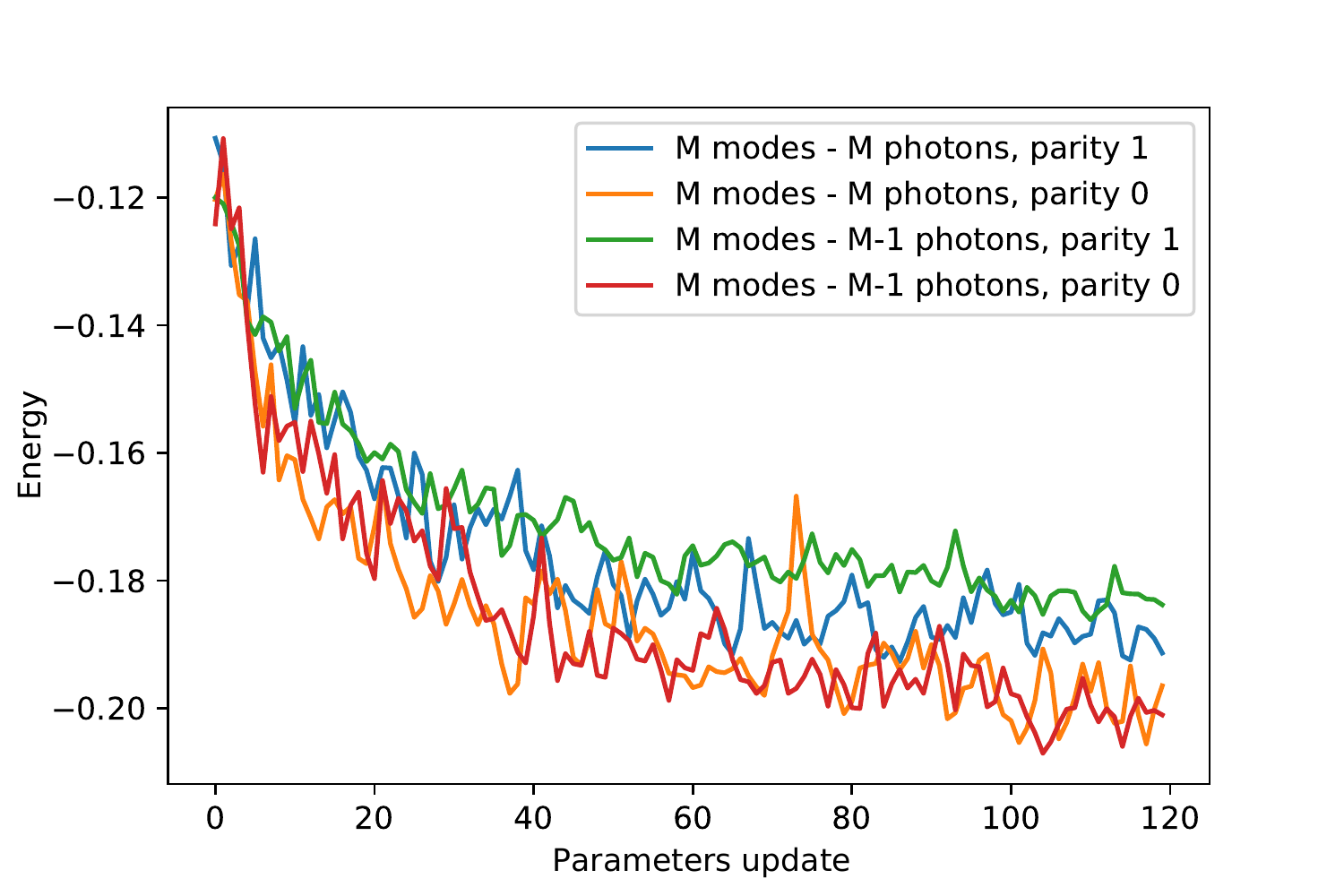}
	\end{subfigure}%
	\caption{Portfolio optimization for 20 companies, $N_q=3$ and $\gamma=1$. (Left) Approach~1 (QUBO). This formulation consists of
    adding a penalty to the bit strings that do not meet the constraint $\sum_i\om_i=1$. The high energies of the first iterations are due to the fact that most generated bit strings  are not an acceptable solution. The best found  solution is $E=-0.2109$ (Sharpe ratio of $4.51$). (Right) Approach~2 (non-QUBO). Each generated candidate solution  is normalized to satisfy the condition $\sum_i\om_i=1$. The found minimum  is $E_{min}=-0.2217$ (much better Sharpe ratio of 6.68 compared to the QUBO formulation).}
\label{fig:approach}
\end{figure}
\begin{figure}[t]
	\includegraphics[width=\textwidth]{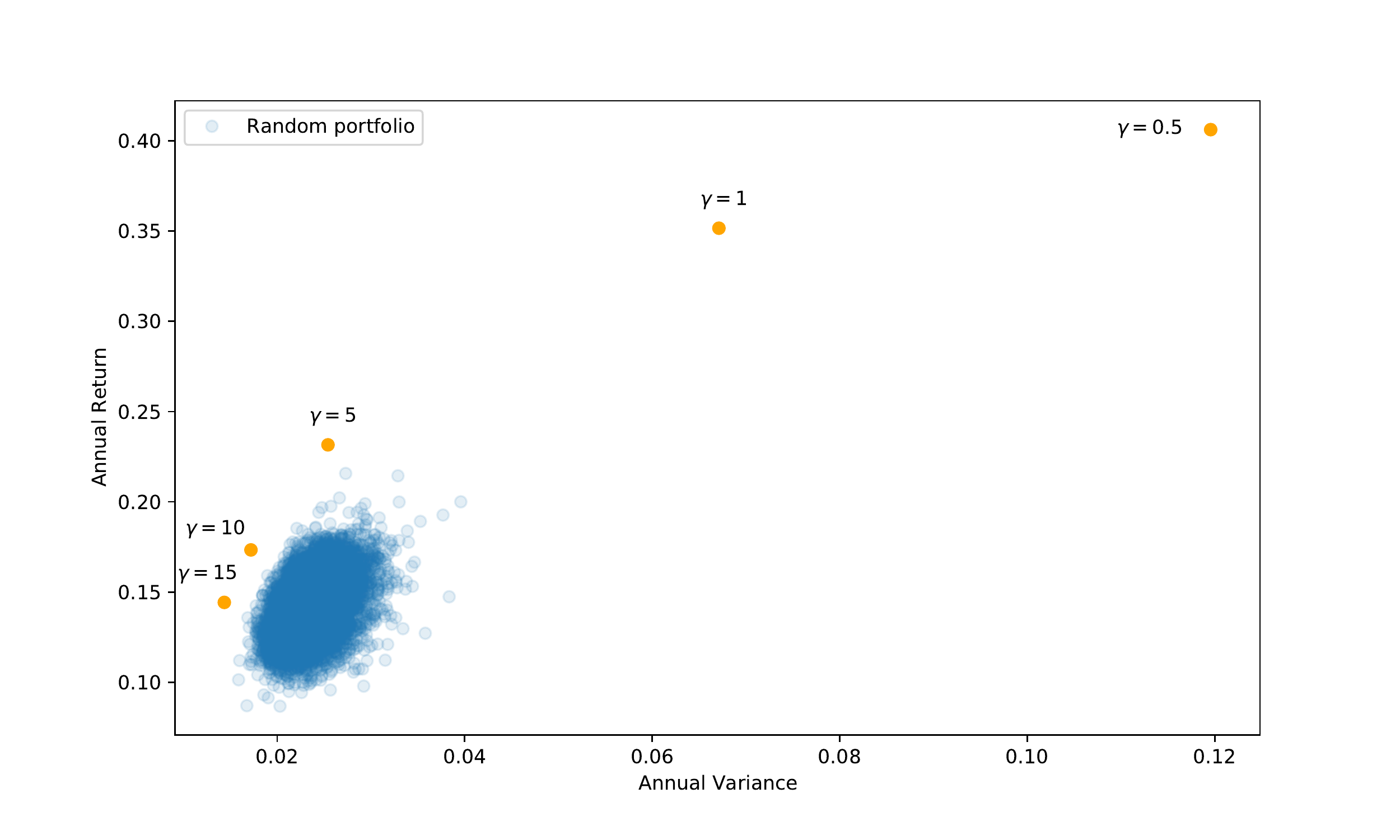}
	\caption{Portfolio of 40 companies where the investment is binary where we decide to invest or not in a company. There are $2^{40} \simeq
    10^{12}$ possible configurations. The $x$-axis corresponds to the risk whereas the $y$-axis is the return of the portfolio. The blue points are 10000 random portfolios. With the same risks, the points found by the bosonic solver (orange points following the efficient frontier) lead to better returns. We present solutions corresponding to different risk aversion~$\g$. }
	\label{fig:portfolio}
\end{figure}
We now consider the  problem of a static portfolio with binary investment, where for each company we decide to invest or not. It is known to be an intractable problem~\cite{venturelli2019reverse}. We formulate it as the second (non-QUBO) approach by optimizing Eq.~\eqref{eq:hamiltonianPortfoliononlinear} for $N_q=1$. The data of 40 companies was taken from Yahoo finance. The algorithm is explained in more detail on p.~\pageref{alg:2}. In Fig.~\ref{fig:portfolio}, we have plotted the solutions found by the variational bosonic solver on a return--risk graph for different risk aversions $\gamma$. The blue points correspond to 10000 random portfolio configurations. We can see that the orange points have the shape of the usual efficient frontier of portfolio optimisation. Also, the random points do not have as  good returns as the orange solutions for a given risk.

\subsection{Improvement by increasing the circuit depth and adding phases}\label{s:deeperCircuit}
As we have witnessed many times (see Fig. ~\ref{fig:randomQUBO2}), the ansatz generated by the optical circuit  does not lead to a single bit string, even in the case of a non-degenerate solution. This is expected given the probabilistic nature of quantum measurement. Also, our circuits are shallow and the performance can be improved by deepening the circuit, that is, by adding more free parameters and thus increasing the expressivity of the ansatz. 

\begin{figure}[t]
	\begin{subfigure}[t]{13cm}
		\resizebox{14cm}{!}{\includegraphics{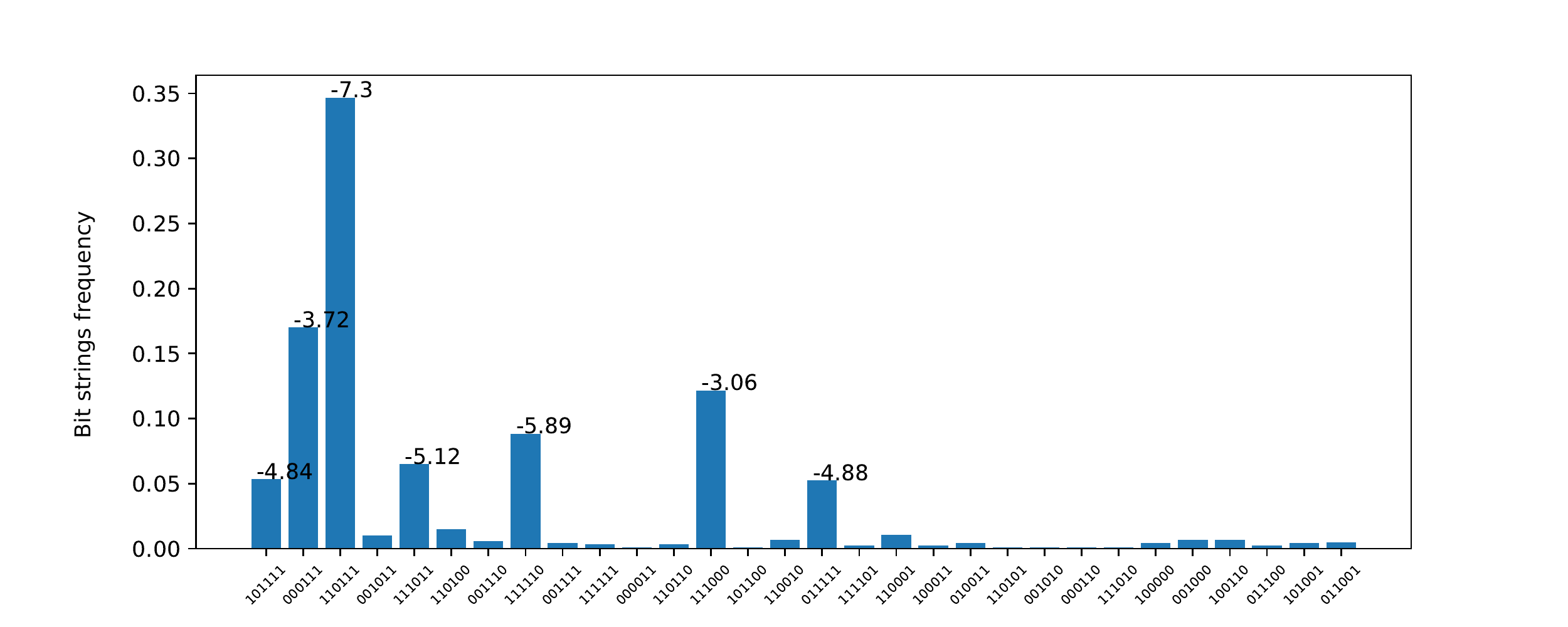}}
	\end{subfigure}
	\begin{subfigure}[t]{14cm}
		\resizebox{14cm}{!}{\includegraphics{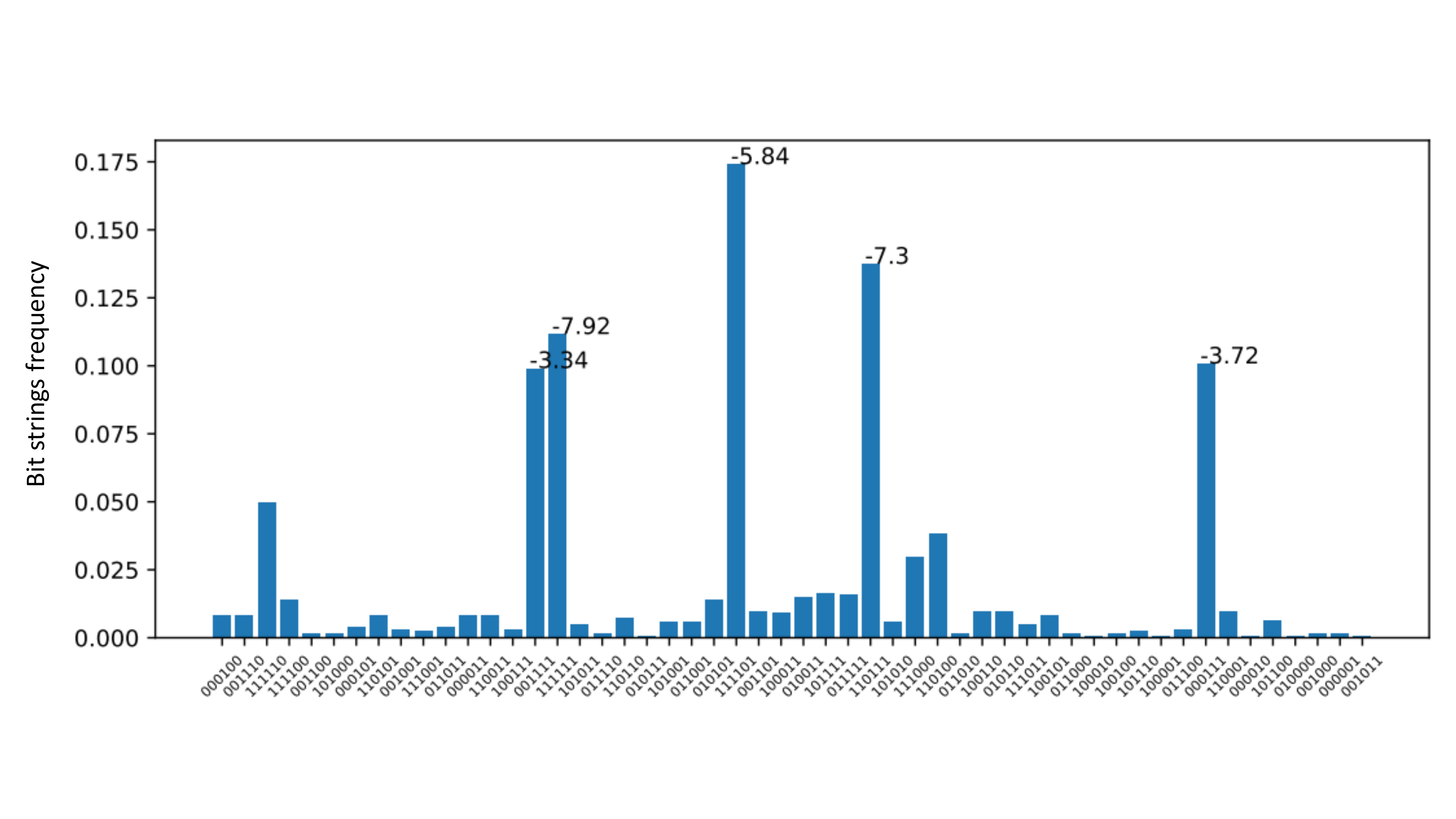}}
	\end{subfigure}
	\caption{Performance improvement for a deeper circuit when solving a QUBO problem defined on a random graph (see Appendix for the actual
    matrix~$Q$). (Top) Sampling from the shallowest circuit $\rC_1(6,n)$. (Bottom) The same problem solved using a deeper circuit allowing us to access the Hilbert space $\rC_2(6,n)$. For both graphs $N_s = 300$ is the number of samples. A brute-force computation shows that the minimum energy is $E_{min}=-7.92$, which is one of the peaks of the bottom graph. The second smallest energy is $E=-7.3$. Thus, the shallowest circuit allows us to find a good approximate solution but a deeper circuit favours the global minimum.}
	\label{fig:2loops}
\end{figure}

Fig.~\ref{fig:2loops} shows how generating a state with a deeper circuit can increase the performance. The plot on the top corresponds to the shallowest optical interferometer sampling from the shallowest space~$\rC_1(8,n)$. The bottom one corresponds to $\rC_2(8,n)$ (first two slices of the Reck scheme). In the first case, the lowest energy we found is $E=-7.3$ corresponding to the bit string $\bb=(1,1,0,1,1,1)$. In the second case of the deeper circuit $\rC_2(8,n)$, the algorithm found the theoretical minimum energy $E_{min}=-7.92$. In Appendix we show the learning curves and the used parameters.

In addition to adding more beam splitters by deepening the circuit, one could also add phase shifters to be optimized during the gradient descent as well, see the black dots in Fig.~\ref{fig:reck}. The phases play no role in the case of the shallowest circuits $\rC_1(M,n)$ but for the deeper ones they could greatly increase the ansatz expressivity.

\subsection{Tolerance to finite sampling}\label{s:finitesampling}
\begin{figure}[t]
	\resizebox{15.5cm}{!}{\includegraphics{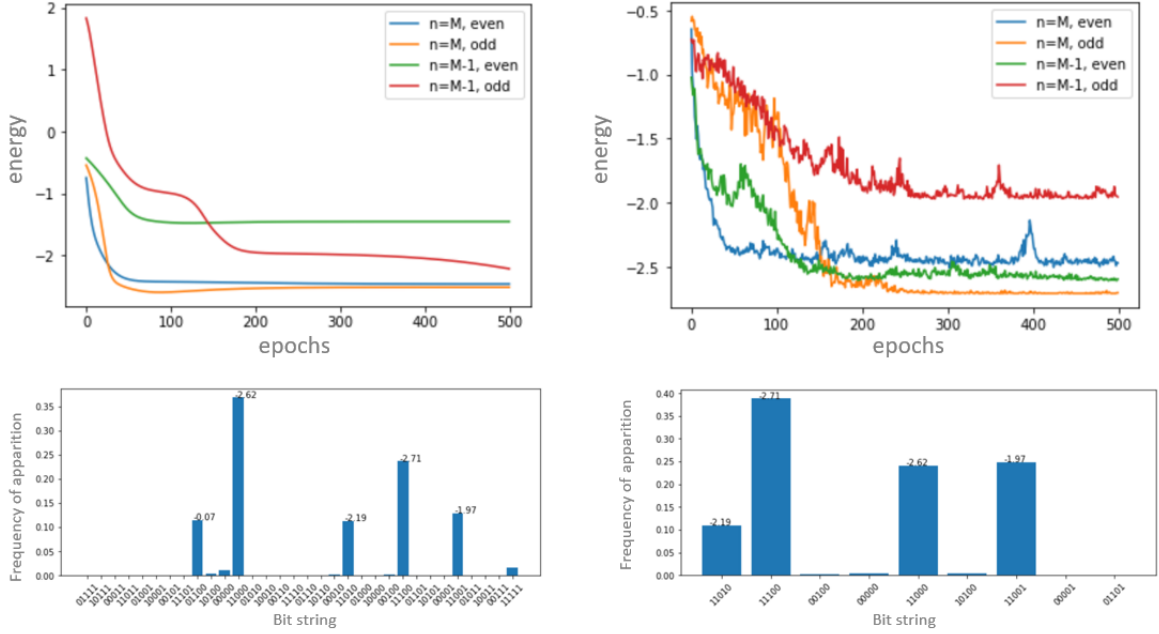}}
	\caption{The impact of finite sampling is illustrated. The figures on the left show the learning curves and solutions for infinite sampling using analytically calculated probabilities. The figures on the right depict the same problem but for finite sampling with $N_s = 400$. We can see that even though the fluctuations are slowing down the convergence, they allow us to explore configurations corresponding to a better solution (optimal in this case). }
	\label{fig:sampling}
\end{figure}

One of the biggest challenges an optimization algorithm faces is being stuck in a local minimum. Quantum variational solvers are not an exemption~\cite{2021arXiv210402955R} and, for example, stochastic gradient descent can sometimes help. Here we observe a similar effect: the fluctuations resulting from the finite sampling measurement are creating randomness during the gradient descent which can also allow us to escape a local minimum (see Fig.~\ref{fig:sampling}). We simulate a shallow interferometer~$\rC_1(6,n)$ in order to solve a QUBO problem and we show the learning curves and the final distribution in two simulations: one with an infinite sampling corresponding to analytically calculating the complete output quantum state and all its probabilities (left) and one with a finite sampling for $N_s = 400$. We  observe that the learning curves are much more fluctuating in the case of finite sampling. It turns out that the fluctuations lead to the optimal solution (the yellow curve on the right).

\subsection{Practical implementation of (shallow) circuits $\rC_i(M,n)$}\label{s:TBI}
\begin{figure}[h]
	\resizebox{13cm}{!}{\includegraphics{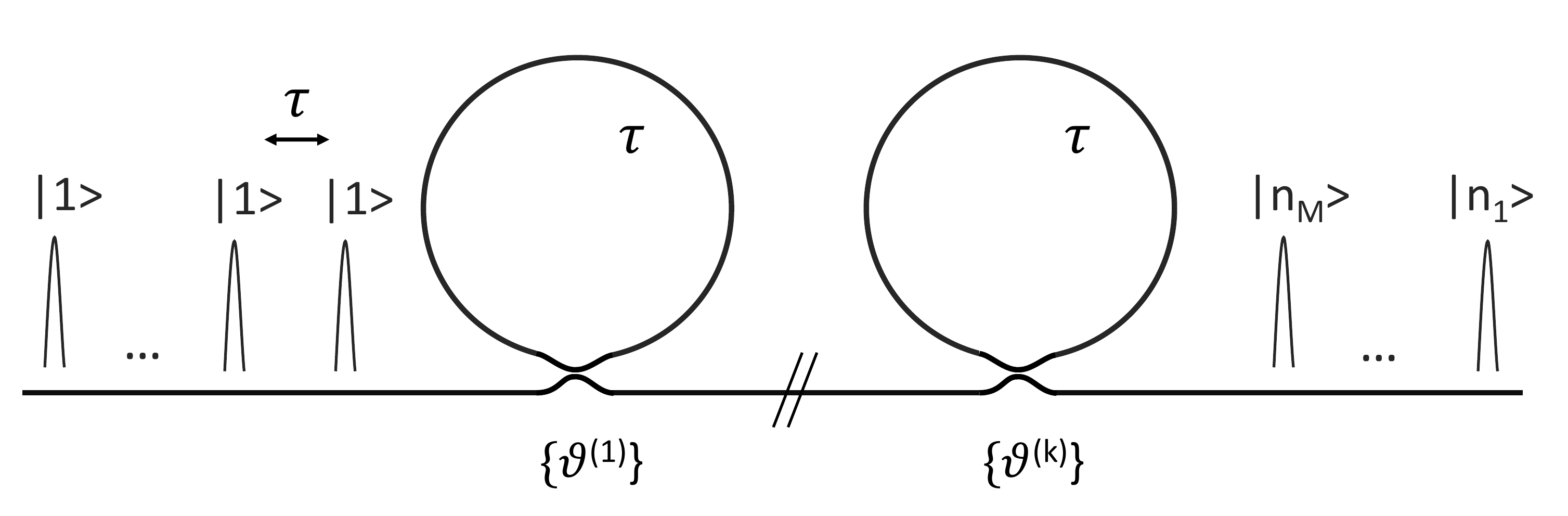}}
	\caption{Experimentally simple implementation of a variational bosonic solver. The input single photons are sent time-separated by $\tau$, which
    also corresponds to the loops' length. The photons interfere at a beam-splitter with a variable angle $\vt_i$. The $i$-th loop corresponds to the $i$-th slice of Reck's scheme. The shallow Hilbert spaces $\rC_i(M,n)$ for $n=M,M-1$ we study in this paper are implemented by $i$ loops and by sending $M$ or $M-1$ consecutive single photons. The variable phase shifters can be inserted in the loops to increase the ansatz expressivity.}
	\label{f:TBI}
\end{figure}
So far we have been implementation-agnostic in our analysis. The family of shallow circuits (and the corresponding Hilbert spaces that is accessed) can be based in bulk optics, optical integrated chip or just any bosonic system with possibly a different definition of a shallow circuit given by a different circuit geometry~\cite{clements2016optimal}. As we already mentioned, the $\rC_i(M,n)$ spaces can be seen as the slices of Reck's scheme~\cite{reck1994experimental} and that, on the other hand, can be conveniently implemented as a multi-loop time-bin interferometer, see Fig.~\ref{f:TBI}. In principle, $M-1$ loops can address the whole group $U(M)$ but in practice a deep circuit like this suffers from exponential photon loss. We aim at a fixed depth (number of loops) and increasing the number of photons modes $M$ and therefore photon number $n=M$ and $n=M-1$. Our theoretical results in Sec.~\ref{s:techdetails} show that even the shallowest circuit of Reck's type can access the qubit Hilbert space of size~$M$.

\subsection{Scalability and other challenges}\label{s:challenges}

How plausible is it to scale the presented variational method? The first problem the reader can point out is the fact that despite the parity map effectively coarse graining  the measurement results of an $M$-mode circuit it is not a sustainable strategy since the size of an $M$-qubit Hilbert space grows exponentially. Recall that the size of any shallow circuit $\rC_i(M,n)$ grows exponentially with $M$ for all $i$. As we argue in detail in Sec.~\ref{s:parityFcn}, the parity functions $\wp_j$ map the basis spanning  $\rC_i(M,n)$ to the basis spanning the $M$-qubit Hilbert space. This space obviously grows exponentially as well but nearly not as fast as $\rC_i(M,n)$ (see the ratio for $i=1$ in Eq.~\eqref{eq:ratioCat} investigated for a different purpose). But here we tacitly assume that the measurement outcomes are uniform which is not the case. We observed that randomly chosen parameters of the studied shallow circuits typically result in a small set of measurement patterns to have high probability enabling it to be sampled with confidence with a bounded number of repeated measurements. The caveat is, of course, that even if this trend continues as $M$ grows it inevitably means that the ratio of such reliably estimated measurements with respect to all possible patterns decreases exponentially. On the one hand, this still enables us to use the proposed variational algorithm. However, the odds of getting stuck in a local minimum most likely increase. How exactly it affects the ability to reach a global solution is a matter of a more detailed study.

Another issue we haven't studied in detail is the effect of photon loss~\cite{brod2020classical,oszmaniec2018classical,moylett2018quantum}. This is the dominant error mechanism in photonic platforms for quantum computing. The theoretical results we present here are valid for a photonic circuits of any depth (from $\rC_1(M,n)$ to the full depth space $\rH^+(M,n)$). Nevertheless, we are interested in shallow circuits $\rC_i(M,n)$ for a fixed small $i$ and arbitrary size~$M$. Hence the effect of photon loss is limited. A particularly convenient way of implementing such circuits is by encoding information in the time of the arrival of the photons (temporal encoding). In that case we need only a few simple optical components, where $i$ is simply the total number of beam-splitters and loops. We leave the question of what how shallow a circuit has to be, where we can expect any quantum advantage, unanswered. The majority of simulations done in this paper are for $\rC_1(M,n)$ where it is know that it can be simulated efficiently classically~\cite{lubasch2018tensor}.

\section{Technical details}\label{s:techdetails}

\subsection{Mapping of a Bosonic Hilbert space basis to a qubit Hilbert space basis}\label{s:parityFcn}

To uncover the behavior of~\eqref{eq:parityFcn} we find useful to introduce different types of integer decompositions. Perhaps the best known is the integer partition and it will be properly introduced in Sec.~\ref{s:ferrers}. For our immediate needs we mention a \emph{{$k$-composition}} and a \emph{weak $k$-composition}~\cite{stanley1997enumerative}. A $k$-composition is a partition of an integer~$n$ into $k$ positive parts where the order matters. The number of compositions of $n$ into exactly $k$ parts is equal to the binomial coefficient $\binom{n-1}{k-1}$. Its weak version is a partition of $n$ into $k$ non-negative parts where the order again matters. It is given by $N(n,k)=\binom{n+k-1}{n}$ and it does not come as a surprise that it coincides with $|\rH^+(M,n)|$ for $k=M$.

Despite the simplicity of the parity function, Eq.~\eqref{eq:parityFcn}, the collective behavior when acting on $\n=(n_1,\dots,n_M)$ is unexpectedly rich. We will study four different situations: $M$ and $n$, even or odd, with a particular emphasis on the physically relevant situation $n\leq M$ mentioned earlier in the main text.
\subsubsection*{$M$ even and $n$ even}
Recall that $n=\sum_{i=1}^{M}n_i$. We introduce a canonical (or representative) form of a measurement pattern such that even numbers are followed by odd numbers:
\begin{equation}\label{eq:nCanonMnEven}
  \n\sim(\underbrace{e,\dots,e}_m,\underbrace{o,\dots,o}_{M-m}),
\end{equation}
where $e,o$ stands for even/odd and take care of all permutations. If both $M,n$ are even then $m=0,2,\dots,M$.  We define  even and odd sums of the canonical form as follows:
\begin{subequations}\label{eq:evenOddSums}
\begin{align}
 \s_e &={1\over2}\sum_{i=1}^{m}n_i, \\
 \s_o &={1\over2} \big(\sum_{i=m+1}^{M}n_i-(M-m)\big)\\
      &={1\over2}\big(n-\sum_{i=1}^{m}n_i-(M-m)\big)\\
      &={1\over2}(n-2\s_e-(M-m))\label{eqs:sigo},
\end{align}
\end{subequations}
where in the first row we transformed each even $n_i$ to $n_i/2$ to transform them to integers. Similarly, for each odd number we summed over $n_i\to (n_i-1)/2$ in the second row and expressed it with the help of the even sum~$\s_e$ in the last row. The transformation guides us to properly use the weak $m$-composition of $\s_e$ for the even part of the canonical detection pattern and the weak $(M-m)$-composition of $\s_o$ for the odd part:
\begin{subequations}\label{eq:ups0}
  \begin{align}
   \upsilon_0(M,n,m) & \df \sum_{\s_e,\s_o} N(\s_e,m)N(\s_o,M-m)\\
     & = \sum_{\s_e,\s_o}\binom{\s_e+m-1}{\s_e}\binom{\s_o+(M-m)-1}{\s_o} \\
     & = \sum_{\s_e=0}^{{1\over2}(n-M+m)}\binom{\s_e+m-1}{\s_e}\binom{{1\over2}(n-2\s_e+M-m)-1}{{1\over2}(n-2\s_e-M+m)}\\
     & = \binom{{1\over2}(n+M+m)-1}{{1\over2}(n-M+m)},\label{eq:multiplicityEE0}
  \end{align}
\end{subequations}
where in the second row we used~\eqref{eqs:sigo} and in the third row we used the identity~\cite{bradler2010conjugate}
\begin{equation}\label{eq:binomIdentity}
  \sum_{s=0}^{r}\binom{p+s}{s}\binom{q-s}{r-s}=\binom{p+q+1}{r}.
\end{equation}
We can learn a few interesting facts from~\eqref{eq:multiplicityEE0}. By counting the number of states of the form~\eqref{eq:nCanonMnEven} we found how many bosonic states are mapped by virtue of~$\wp_0$ to a qubit basis state with the first~$m$ zeros followed by $M-m$ ones
\begin{equation}\label{eq:wp0action}
  \wp_0\colon(\underbrace{e,\dots,e}_m,\underbrace{o,\dots,o}_{M-m}) \mapsto(\underbrace{0,\dots,0}_m,\underbrace{1,\dots,1}_{M-m}).
\end{equation}
Beside the `degeneracy' of each boson basis state we also know  how big the qubit Hilbert space~$\H_{M}$ is. To this end, let's set $n=M$. This corresponds to the biggest bosonic Hilbert space $\rH^+(M,M)$ an $M$-mode linear interferometer is able to address (a full-depth circuit). We can see that $\upsilon_0(M,M,m)$ is nonzero for all even~$m$'s. The same conclusion holds for all $\binom{M}{m}$ permutations of~\eqref{eq:nCanonMnEven}. Since $\sum_{m=0,2,\dots}^{M}\binom{M}{m}=2^{M-1}$ it follows that $\wp_0$ is a surjective function from the set of basis states of $\rH^+(M,M)$ to the set of basis states of $\H_{M-1}$.

Since only bit strings with an even number of zeros and ones are in the range of $\wp_0$ we may need to rename the basis states (for example, by ignoring the least significant bit). A better solution with a bigger qubit Hilbert space will emerge once we study the case of $n$ odd. As an additional comment, thanks to~\eqref{eq:multiplicityEE0}, once we know the probability distribution in the bosonic Hilbert space for each basis state mapped to a given $(M-1)$-qubit state (where some argue that it is an intractable task~\cite{aaronson2011computational}) we know the distribution in the qubit Hilbert space.

The action of the second parity function $\wp_1$ differs only by swapping the bit positions:
\begin{equation}\label{eq:wp1action}
  \wp_1\colon(\underbrace{e,\dots,e}_m,\underbrace{o,\dots,o}_{M-m}) \mapsto(\underbrace{1,\dots,1}_m,\underbrace{0,\dots,0}_{M-m}).
\end{equation}
Hence
\begin{equation}\label{eq:multiplicityEE1}
  \upsilon'_0(M,n,m)= \binom{{1\over2}(n+2M-m)-1}{{1\over2}(n-m)}
\end{equation}
follows from~\eqref{eq:multiplicityEE0} upon transforming $m\to M-m$. Have we achieved something by considering $\wp_1$ compared to $\wp_0$? Thanks to $m,M$ being even $\wp_1$ maps the bosonic state to the same qubit Hilbert space as $\wp_0$ but since $\upsilon_0(M,n,m)\neq\upsilon'_0(M,n,m)$ it is not the same mapping. From the practical perspective it may be advantageous to use $\wp_0$ or $\wp_1$ or both at the same time.

The remaining cases share the derivation details with this case and so we will focus mostly on the different behavior and consequences of the parity functions.

\subsubsection*{$M$ even and $n$ odd}
If $M$ is even and $n$ odd then the number of even and odd digits must be odd.  The canonical detection pattern then looks like
\begin{equation}\label{eq:nCanonMevennodd}
  \n\sim(\underbrace{e,\dots,e}_{m},\underbrace{o,\dots,o}_{M-m}),
\end{equation}
where $e,o$ stand for even/odd and $m=1,3,\dots,M-1$. The same derivation leads us to the identical result for $\upsilon_j$. Hence, Eqs.~\eqref{eq:multiplicityEE0} and~\eqref{eq:multiplicityEE1} hold for $M$ even and $n,m$ both either even or odd. Let's turn our attention to the case $n=M-1$. Inspecting~\eqref{eq:nCanonMevennodd}, we see that the functions $\wp_j$ are again surjective functions whose range is the set of basis state of Hilbert space $\H_{M-1}$. This is because $\sum_{m=1,3,\dots}^{M-1}\binom{M}{m}=2^{M-1}$ followed by the same process of erasing the least significant bit of the basis bit string. But this Hilbert space is orthogonal to the one obtained for $n$ even. Their union is the basis set of the $2^M$-dimensional Hilbert space $\H_M$ and so in order to access it when $M$ is even we run two experiments: one with $n=M$ and another one with $n=M-1$.

\subsubsection*{$M$ odd and $n$ even}
The parity function $\wp_j$ acts quite differently when $M$ is odd. This can be ultimately tracked down to
\begin{equation}\label{eq:nCanonModdneven}
  \n\sim(\underbrace{e,\dots,e}_{m~\mbox{\scriptsize odd}},\underbrace{o,\dots,o}_{M-m~\mbox{\scriptsize even}}),
\end{equation}
where $m=1,3,\dots,M-1$. The parity function acts like in~Eqs.~\eqref{eq:wp0action} and~\eqref{eq:wp1action} but unlike the two previous cases, where both $m$ and $M-m$ were either even or odd, now they have an opposite parity. The effect is that their ranges are disjoint and if we set $n=M-1$ as our physically well-motivated setup the union of the disjoint ranges is nothing else than the full basis set of~$\H_M$. Indeed, by repeating the derivation of $\upsilon_0(M,n,m)$ and $\upsilon'_0(M,n,m)$ we get the same result, Eqs.~\eqref{eq:multiplicityEE0} and~\eqref{eq:multiplicityEE1}, further extending its validity. They are non-zero for all $m=1,3,\dots,M-1$ and therefore for all the permutations. For $n=M-1$ and by using the orthogonality argument together with $2\sum_{m=1,3,\dots}^{M-1}\binom{M}{m}=2\times2^{M-1}=2^M$ the claim follows.

\subsubsection*{$M$ odd and $n$ odd}
Finally, when both $M,n$ are odd, that is
\begin{equation}\label{eq:nCanonModdnodd}
  \n\sim(\underbrace{e,\dots,e}_{m~\mbox{\scriptsize even}},\underbrace{o,\dots,o}_{M-m~\mbox{\scriptsize odd}}),
\end{equation}
where $m=0,2,\dots,M-1$, we arrive at the same conclusion by following the same arguments in the previous case. Hence, Eqs.~\eqref{eq:multiplicityEE0} and~\eqref{eq:multiplicityEE1} are valid for any $M,n$ and all admissible $m$'s. If we set $n=M$ then $\upsilon_0(M,M,m)$ and $\upsilon'_0(M,n,m)$ are nonzero for all $m$ even and  the union of the disjoint ranges of $\wp_0$ and $\wp_1$ is the $2^M$ canonical bases of $\H_M$ thanks to $2\sum_{m=0,2,\dots}^{M-1}\binom{M}{m}=2^M$. Even thought the Hilbert spaces are the same for $n$ even or odd the parity function assign different weights to different qubit bases. To democratize the Hilbert space it again seems useful to run two experiment for $n=M$ and $n=M-1$ exactly like for $M$ even. Note, however, that the reason the $2^M$-dimensional Hilbert space $\H_M$ `comes together' is quite different.

We have obtained a complete understanding of how the parity function $\wp_j$ acts on the basis of  $\rH^+(M,n)$ but this is not the Hilbert space we can easily access for any~$M$ due to its high-depth. As indicated in the main text, we are  interested in much shallower circuits that are related to $\rH^+(M,n)$ in many ways as we will see. In order to get a proper handle on these we will introduce several concepts from discrete mathematics and the representation theory of the symmetric group~$\mathrm{S}_n$.

\subsection{Lattice paths and Catalan numbers}\label{s:latticePath}
\begin{figure}[t]
  \resizebox{13cm}{!}{\includegraphics{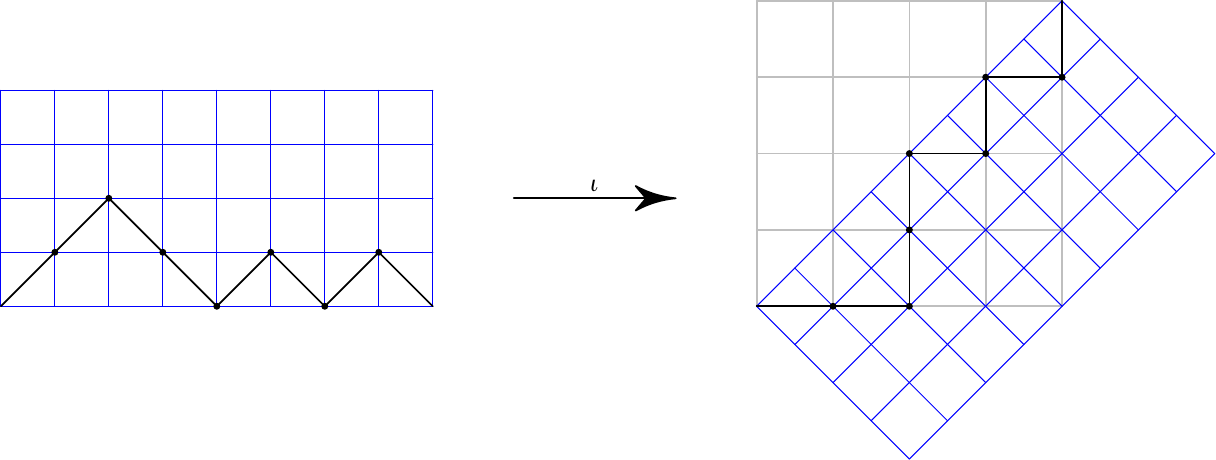}}
  \caption{A Dyck path from the set $\euD(8,0,0)$ generated by the Dyck word $UUDDUDUD$ on the left and the same path mapped by~$\iota$ defined in~\eqref{eq:isoDyck} on the right. The grey lattice is where the Dyck path can be seen as a staircase path. }
  \label{fig:dyck}
\end{figure}

Consider the square lattice  in the first quadrant  $Q_{++}=\bbN\times\bbN$, where $\bbN$ are non-negative integers, and let $p_j=(j,y_j)\in Q_{++}$ for $j=[0,k]$ be a lattice point such that $|y_{j+1}-y_j|=1$. Then the sequence $(p_0,\dots,p_k)$ is called a \emph{Dyck path}, where the ascending/descending segment $(p_j,p_{j+1})$ satisfies $y_{j+1}-y_j=\pm1$. Hence a Dyck path is a lattice path where one travels either north-east ($U$ as UP) or south-east ($D$ as DOWN). Every Dyck path can be written in the form of a \emph{Dyck word} as a $k$-length string of $U$'s and $D$'s. The counting of Dyck paths becomes interesting once we set the initial point $(0,y_0)$ and the terminal point $(k,y_k)$. We will use $y_0=\d_1,y_k=\d_2$ and label the set of all such Dyck paths $\euD(k,\d_1,\d_2)$. It is known~\cite{feller2008introduction} that
\begin{equation}\label{eq:CardDyckPaths}
  |\euD(k,\d_1,\d_2)|=\binom{k}{{1\over2}(k+\d_2-\d_1)}-\binom{k}{{1\over2}(k-\d_2-\d_1-2)}.
\end{equation}
We see an example in the left panel of Fig.~\ref{fig:dyck} for $k=8,\d_1=\d_2=0$. For $\d_1=\d_2=0$ the expression becomes the $k/2$-th Catalan number~\cite{koshy2008catalan,stanley2015catalan}
\begin{equation}\label{eq:Catalan}
  C_{k/2}={2\over2+k}\binom{k}{{k\over2}}.
\end{equation}
We introduce a bijective mapping of the lattice (and therefore all its paths):
\begin{equation}\label{eq:isoDyck}
  \iota\df R(-\pi/4)\circ R_{x},
\end{equation}
whose action is in the right panel of Fig.~\ref{fig:dyck}. This is one of many incarnations of Dyck paths, Catalan numbers and their generalizations. In this case one can introduce a different lattice (grey) and the Dyck paths become the so-called staircase paths on the new lattice~\cite{mohanty1979lattice}. For example, the constraint on all Dyck paths with $\d_1=\d_2=0$ dictates that none can `dip' below the $x$ axis of the first quadrant. Under the action of~$\iota$ it becomes the condition forbidding a staircase path to cross the diagonal of the grey lattice. We will give a convenient physical interpretation of the grey lattice for all Dyck paths $\euD(k,\d_1,\d_2)$. Note that the bosons and Dyck paths are two very closely related objects~\cite{bradler2015hiking}.

\subsection{Integer partitions, Ferrers diagrams and Young's lattice}\label{s:ferrers}

\begin{figure}[t]
  \resizebox{8cm}{!}{\includegraphics{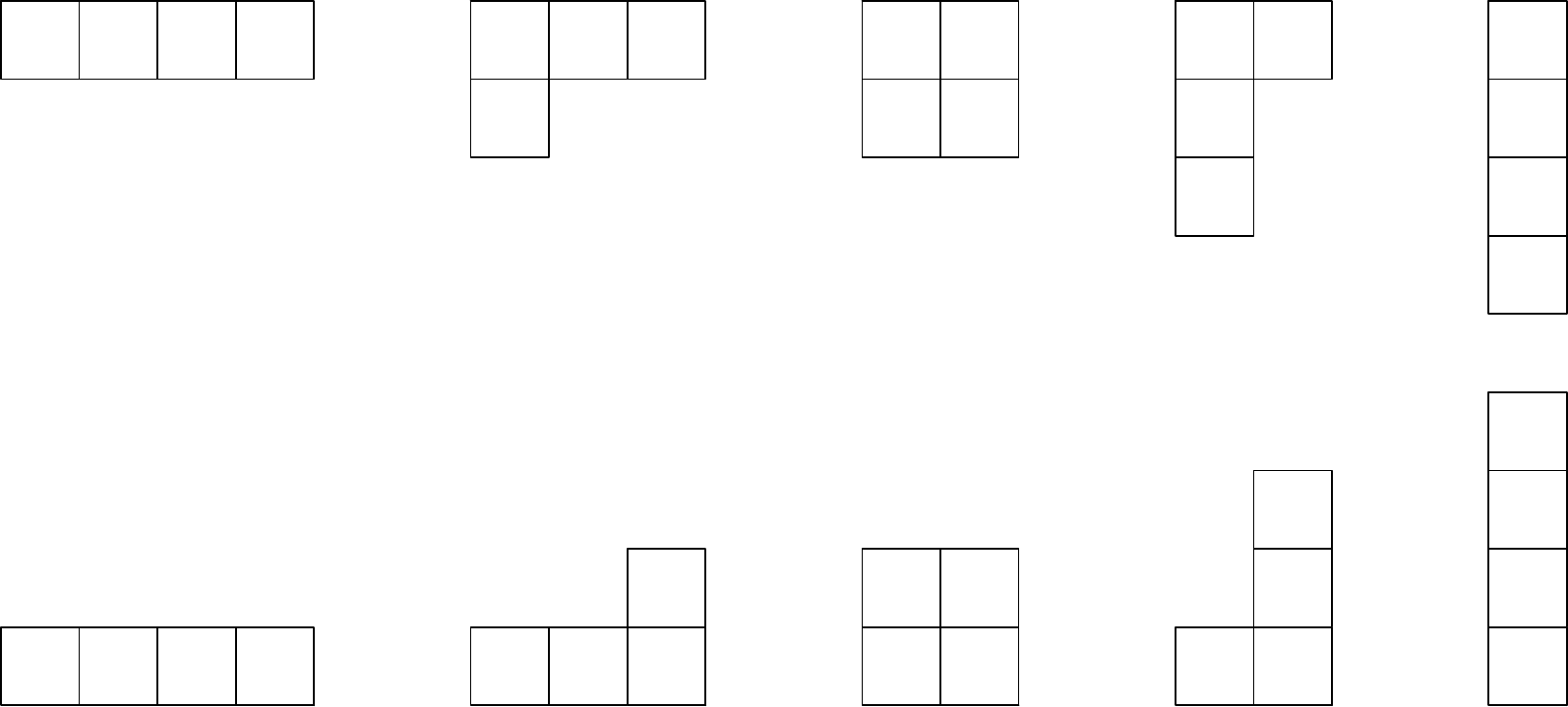}}
  \caption{(Top) The English convention for Ferrers (Young) diagrams showing the five integer partitions $\la\vdash n$ for $n=4$: $(4), (3,1), (2,2), (2,1,1),(1,1,1,1)$. (Bottom) The reflected French convention for partitions of~$n=4$ used throughout this paper: $(1,1,1,1),(1,1,2),(2,2),(1,3)$ and $(4)$.}
  \label{fig:F1}
\end{figure}

A partition of an integer $n>0$ is a sequence of positive integers $\la=(\la_1,\dots,\la_k)$ satisfying $\la_1\geq\la_2\geq\cdots\geq\la_k>0$ such that $\sum_{i=1}^{k}\la_i=n$. For a partition of $n$ by $\la$ we write $\la\vdash n$. A Ferrers (or Young) diagram is a graphical notation for a partition of an integer~$n$~\cite{andrews2004integer}. It consists of~$n$ boxes arranged in $k$ rows, where the $i$-th one contains $\la_i$ of them (the English convention). We will use a different, physically motivated, convention, where for $\la\vdash n$ we write $0<\la_1\leq\la_2\leq\cdots\leq\la_k$. Hence, the Ferrers diagram used here will contain~$n$ boxes, where the $i$-th column is $\la_i$ boxes  stacked on top of each other (the reflected French convention). Both conventions are depicted in Fig.~\ref{fig:F1} for $n=4$.

One can introduce various orderings on integer partitions to promote them to a \emph{poset} (partially ordered set). A poset $P=(Z,R)$ is a set~$Z$ with a partial equivalence relation~$R$ called~a~\emph{partial order}~\cite{stanley1997enumerative}. The partial order on integer partitions we are interested in is called the \emph{inclusion relation} $\la\subseteq\mu$ if $\la_i\leq\mu_i,\forall i$. We then call $\la$ and $\mu$ \emph{comparable} and using the equivalence of integer partitions and Ferrers diagrams we see that two Ferrers diagrams are comparable if one is included in the other. If $\la\subseteq\mu$ and $1+\sum_i\la_i=\sum_i\mu_i$ (that is, one box removed from a Ferrers diagram to get another Ferrers diagram) we write $\la\nearrow\mu$. A~convenient graphical representation of a poset is called a \emph{Hasse diagram}~\cite{stanley1997enumerative}. It is a directed graph, where the partial order is indicated by directed edges. The inclusion poset of integer partitions is special. It has an additional structure promoting it to a~\emph{lattice}~\cite{stanley1997enumerative}. In a lattice, all pairs of set elements have a greatest lower bound (the~\emph{join}) and a smallest upper bound (the~\emph{meet})\footnote{Despite the name, the integer lattice~$Q_{++}$ introduced in Sec.~\ref{s:latticePath} is not a lattice but rather a meet-semilattice (still a poset).}. The set of all partitions of all integers equipped with inclusion is known as \emph{Young's lattice}~$Y$~\cite{sagan2013symmetric}. Pictorially, the join for $Y$ is the intersection of the Ferrers diagrams and the meet is their union. Normally, Young's lattice expands indefinitely in one direction. For us, however, only finite Young's sublattices are physically relevant. Let $\mu$ be any partition. Then, $Y_\mu=(\mu,\subseteq)$ is the lattice of all partitions contained in $\mu$.

\begin{figure}[t]
  \resizebox{15.3cm}{!}{\includegraphics{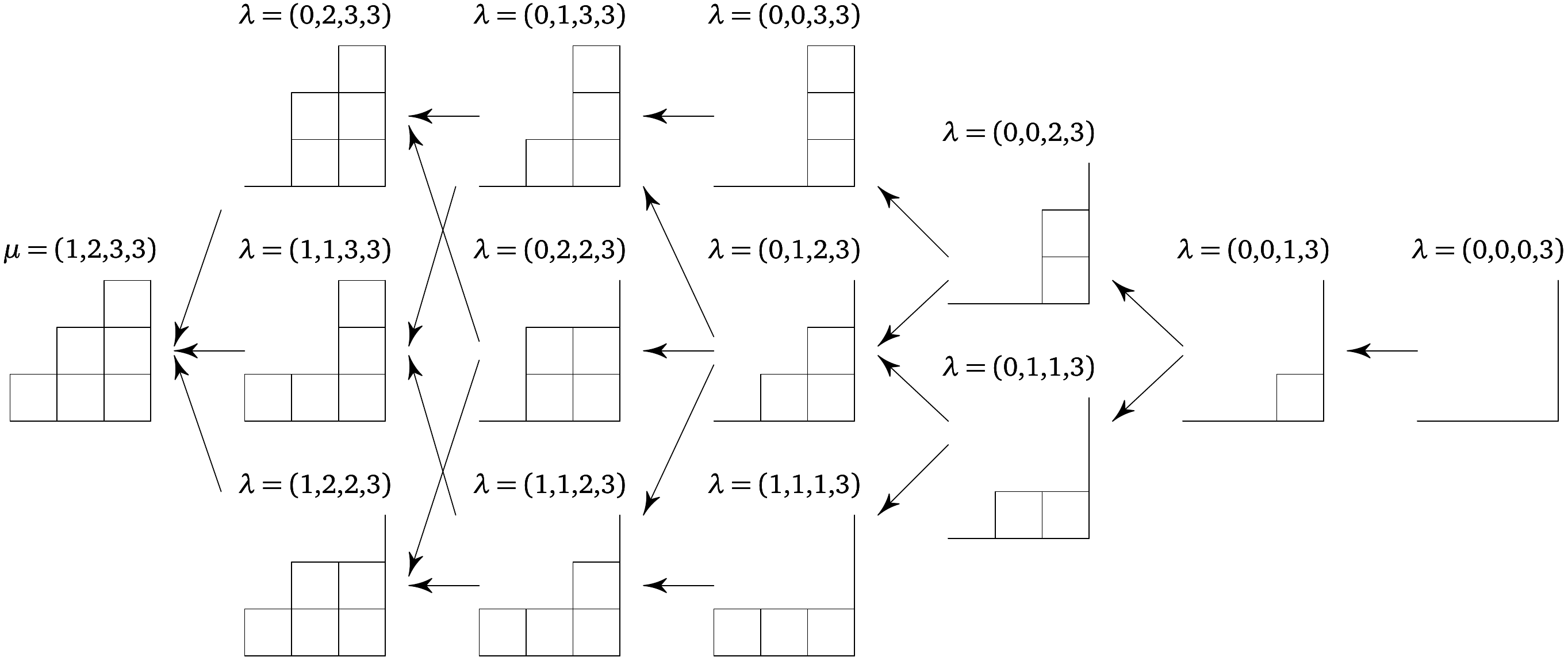}}
  \caption{Young's lattice $Y_{(1,2,3)}$ with the depicted extended Ferrers diagrams. The greatest element $\mu=(1,2,3,3)$ is on the left  and the arrows point in the direction of included diagrams~$\la\nearrow\mu$.}
  \label{fig:F2}
\end{figure}

In order to make contact with the lattice paths we find useful to introduce a slightly redundant notation for  Ferrers diagrams than by counting the boxes in each column. Assume Young's lattice~$Y_\mu$. Instead of the boxes of the included Ferrers diagram we will count the length of all vertical lines in the Ferrers diagram and any $\la\subseteq\mu$ will be created not by removing the boxes but rather by removing the left vertical and upper horizontal segments of a box. It implies a notational change for $\mu$. Contrary to the integer partition, it will always be a sequence of $k+1$ integers  $\la=(\la_1,\dots,\la_k,\la_{k+1})$, where $0\leq\la_1\leq\la_2\leq\cdots\leq\la_{k+1}$, that is, including zeros which are normally omitted for integer partitions. Pictorially, it is as if we were counting the segments of all vertical lines of the columns in the Ferrers diagram (including the removed ones from~$\mu$). Let's call this augmented structure the \emph{extended Ferrers diagram}. We depict Young's lattice $Y_{(1,2,3)}$ but with the extended Ferrers diagram notation and drawings in Fig.~\ref{fig:F2}.

\begin{figure}[h]
  \resizebox{9cm}{!}{\includegraphics{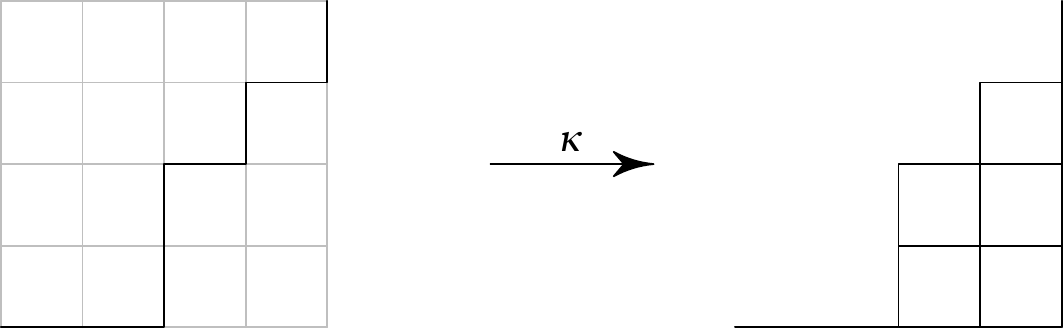}}
  \caption{The bijection $\kappa$ between a staircase lattice path from Fig.~\ref{fig:dyck} connecting the lattice points $(0,0)$ and $(4,4)$  and the extended Ferrers diagram $\la=(0,0,2,3,4)$ corresponding to the partition $(2,3)$ of $n=5$ is depicted.}
  \label{fig:F3}
\end{figure}

The extended Ferrers diagrams can be interpreted as staircase paths starting from the bottom left corner and ending in the top right corner by following the `left perimeter' of the (extended) Ferrers diagram~\cite{mohanty1979lattice}. The isomorphism is illustrated in Fig.~\ref{fig:F3}.
Finally, motivated by the physical circumstances of the studied bosonic systems (see Sec.~\ref{s:shallowCat}) we prepend each $\la$ by $\la_0=0$, thus forming $(\la_0,\la_1,\dots,\la_k,\la_{k+1})$, and take the first differences:
\begin{equation}\label{eq:firstDiffs}
  \n\df(\la_1-\la_0,\la_2-\la_1,\dots,\la_{k+1}-\la_k)=(n_1,\dots,n_{k+1}).
\end{equation}
In other words, $\n$ is the height we have to overcome as we walk on top of the boxes (the extended Ferrers diagrams) from left to right. More importantly, as the notation suggests, this is identified with the detection pattern obtained by the Fock measurements, that first appeared in Sec.~\ref{s:algo}. The reason for introducing Young's lattice and the staircase isomorphism is to explicitly list the Catalan Hilbert space basis whose dimension is provided by the Dyck path isomorphism $\iota$ in~\eqref{eq:isoDyck} and study the properties of the parity function~$\wp_j$. It will help us gain insight into the structure of the output bosonic state. Other, more practical, consequences are discussed in the main text.

\subsection{Shallow bosonic circuits, Catalan Hilbert spaces and Boolean lattices}\label{s:shallowCat}

We will call a subspace of $\rH^+(M,n)$ a Catalan Hilbert space whenever the method of counting the dimensionality involves counting the number of staircase or Dyck paths, see Sec.~\ref{s:latticePath}. This is where the Catalan numbers play a pivotal role. It is purposely a loose definition. For instance, unlike a full-depth circuit, it is not sufficient to specify the number of modes and the total photon number but also the input state  together with the exact mesh description of the beam-splitters' position. So even though the presented method can be used, for practical reasons we investigate Catalan Hilbert space in the well-defined geometries such as the Reck or Clements setup~\cite{reck1994experimental,clements2016optimal}. So, if we decompose a unitary $U$ as maximum-depth optical circuit, we will be interested in the first $n$ layers of evolution operators after suitably `slicing' the unitary $U=\prod_{i=1}^{n}U_i$. This is still possible in more than one way but some are more natural (or useful) than others. For example, in Reck's case we can define a sequence of Catalan Hilbert spaces inclusions
\begin{equation}\label{eq:inclusionCatalan}
\rC_1(M,n)\subsetneq\rC_2(M,n)\subsetneq\cdots\subsetneq\rC_{M-2}(M,n)\subsetneq\rH^+(M,n)
\end{equation}
corresponding to `diagonal' slicing (cf.~Fig.~\ref{fig:reck}). The inclusions manifest as subdiagrams of a Hasse diagram of the Ferrers diagrams (studied in Sec.~\ref{s:ferrers}) corresponding to~$\rH^+(M,n)$. This picture not only sheds light on the structure of the completely symmetric spaces but will also help us understand the parity map restricted to $\rC_i(M,n)$. Recall that we have a good understanding of the behavior of $\wp_j$ for deep circuits from Sec.~\ref{s:parityFcn} due to its simple structure. Also, note that in practice (similarly to $\rH^+(M,n)$) we will focus on $n\leq M$ considering at most one input photon per mode. Before we proceed we present two examples putting together all the introduced concepts so far.
\begin{exa}
  Our first example will be a shallow circuit previously studied in detail~\cite{lubasch2018tensor}, showing that the output can be described by a 1D tensor network state known as a matrix product state. It can be also seen as the first diagonal `slice' of the Reck universal scheme~\cite{reck1994experimental} and it is depicted in Fig.~\ref{fig:circuit1Loop} for four modes (cf. Fig.~\ref{fig:reck}).
  \begin{figure}[h]
  \resizebox{8cm}{!}{\includegraphics{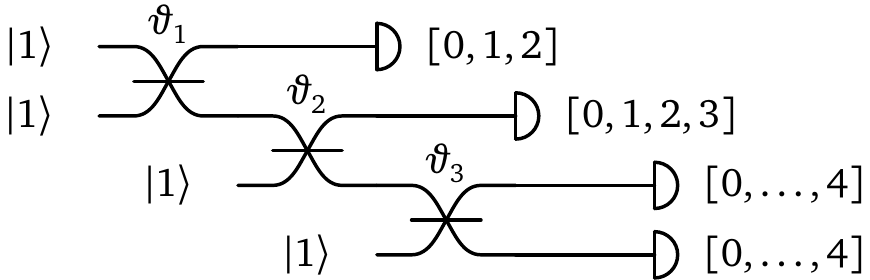}}
  \caption{A one-dimensional `cascade' of tunable beam-splitters parametrized by the angles~$\vt_i$ transforms an input state $\ket{1111}$ into an entangled output state occupying a Catalan Hilbert space whose structure we further investigate. The state is detected by PNRs and the numbers in brackets denote the possible photon measurement outcomes.}
  \label{fig:circuit1Loop}
  \end{figure}
  Motivated by the practical consideration, we are interested in the case of each mode being occupied by a single photon but the method for getting an insight into the output Hilbert space structure based on lattice path counting is applicable more widely. The numbers next to the detectors show the possible photons counts per detector. There are two constraints on the photons counts. First, there is the maximal possible number of detected photons per mode which is the biggest number in the square brackets. Second, the numbers in different brackets are not independent. For example, if we detect two photons in the first detector, we can't possibly detect two photons in the second one (from the top). The both constraints are captured in the left panel of Fig.~\ref{fig:example1Loop}.
  \begin{figure}[h]
  \resizebox{13cm}{!}{\includegraphics{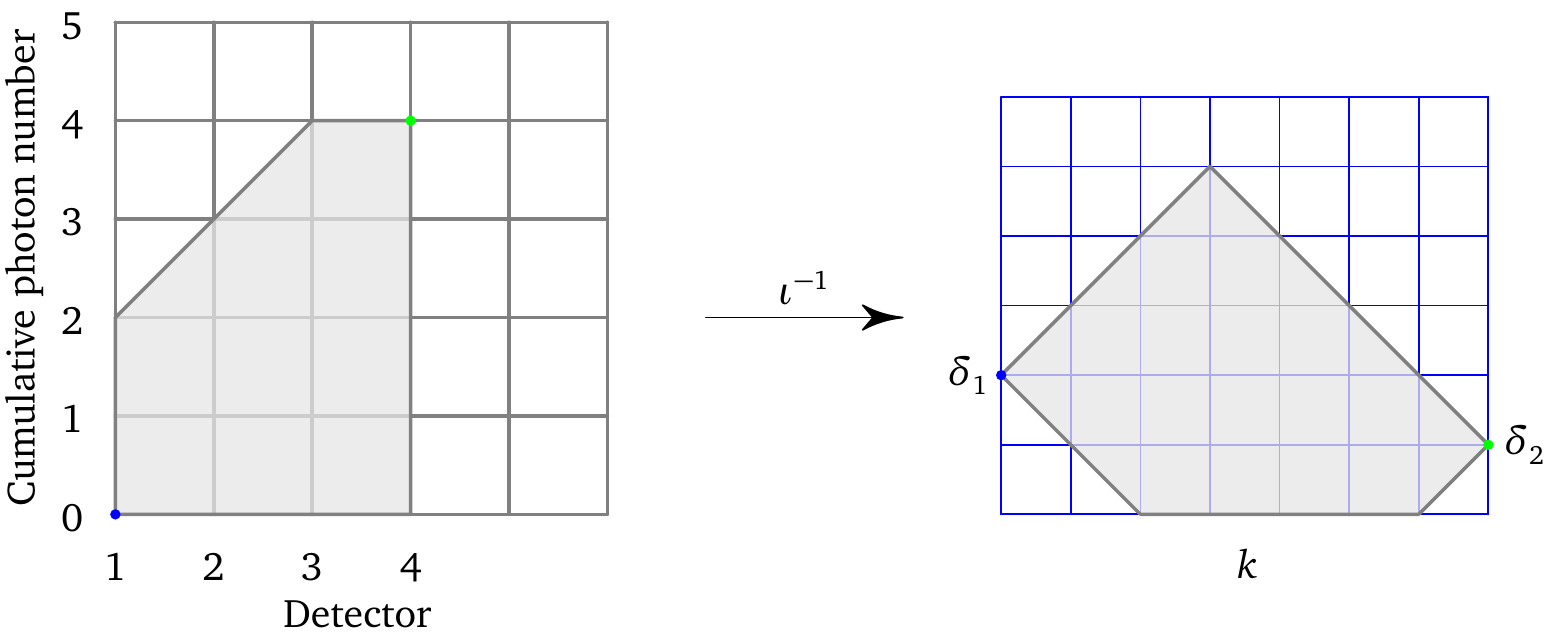}}
  \caption{The shaded polygon in the left plot captures the possible detection patterns of the circuit in~Fig.~\ref{fig:circuit1Loop}. As explained in the main text, all possible detection patterns correspond to the allowed staircase paths in the gray lattice. The detector on the $x$~axis counts the detectors from top to bottom in Fig.~\ref{fig:circuit1Loop}. On the right, we see the same polygon under the inverted action of the isomorphism defined in~\eqref{eq:isoDyck} transforming the staircase paths to Dyck paths in the blue lattice. In both cases, the starting/finishing points are indicated by the blue/green dots. The parameters $k=7,\d_1=2,\d_2=1$ help us count the number of Dyck/staircase paths by virtue of Eq.~\eqref{eq:CardDyckPaths}.}
  \label{fig:example1Loop}
  \end{figure}
  On the $x$~axis of the gray lattice we count the detectors and the $y$ axis is the cumulative detected photon number. The first constraint takes the form of the perimeter of the light-gray (in this case convex) polygon. The second constraint is the cumulative character of the diagram where naturally only the allowed differences of the detected photon numbers are possible. The starting point (before any detection takes place) is the bottom left corner (the blue dot). If the first detector clicks once, we move one segment up and then to the right. If the second detector detects two photons we move two segments up and one to the right. We just hit the boundary and any detection event beyond would be forbidden. We continue and we must end up in the upper right corner (the green dot). This is because the total photon number (4 in this case) is preserved. Every path satisfying the above constraints is allowed and these paths are staircase walks introduced earlier.  We then invert bijection~$\iota$ defined in~\eqref{eq:isoDyck} and depict the same path as a Dyck path in the right panel. This allows us to enumerate the number of possible measurement patterns using Eq.~\eqref{eq:CardDyckPaths}. For $k=7,\d_1=2$ and $\d_2=1$ we find $|\euD(7,2,1)|=28$.

  This number must agree with the second bijection, $\kappa$, between the staircase paths and integer partitions (and therefore the extended Ferrers diagrams) introduced in Sec.~\ref{s:ferrers}. We illustrate it on $Y_{(2,3,5)}$ whose Hasse diagram is in Fig.~\ref{fig:Hasse1}. Each vertex is a Ferrers diagram/staircase path (equivalent under the action of~$\kappa$) and the arrows indicate inclusion (see Sec.~\ref{s:ferrers}). The left plot captures the same information as Fig.~\ref{fig:F2} but we also added two other vertex labels of the same graph, the detection patterns~$\n$ (the middle plot) and the bit string as the result of $\wp_0(\n)$ (the right plot). The principal motivation behind introducing Young's lattices  is to identify certain universally present substructures (namely the Boolean sublattices, see~Sec.~\ref{s:details}) allowing us to formulate some desirable properties of all Catalan Hilbert spaces.
  \begin{figure}[h]
  \resizebox{16cm}{!}{\includegraphics{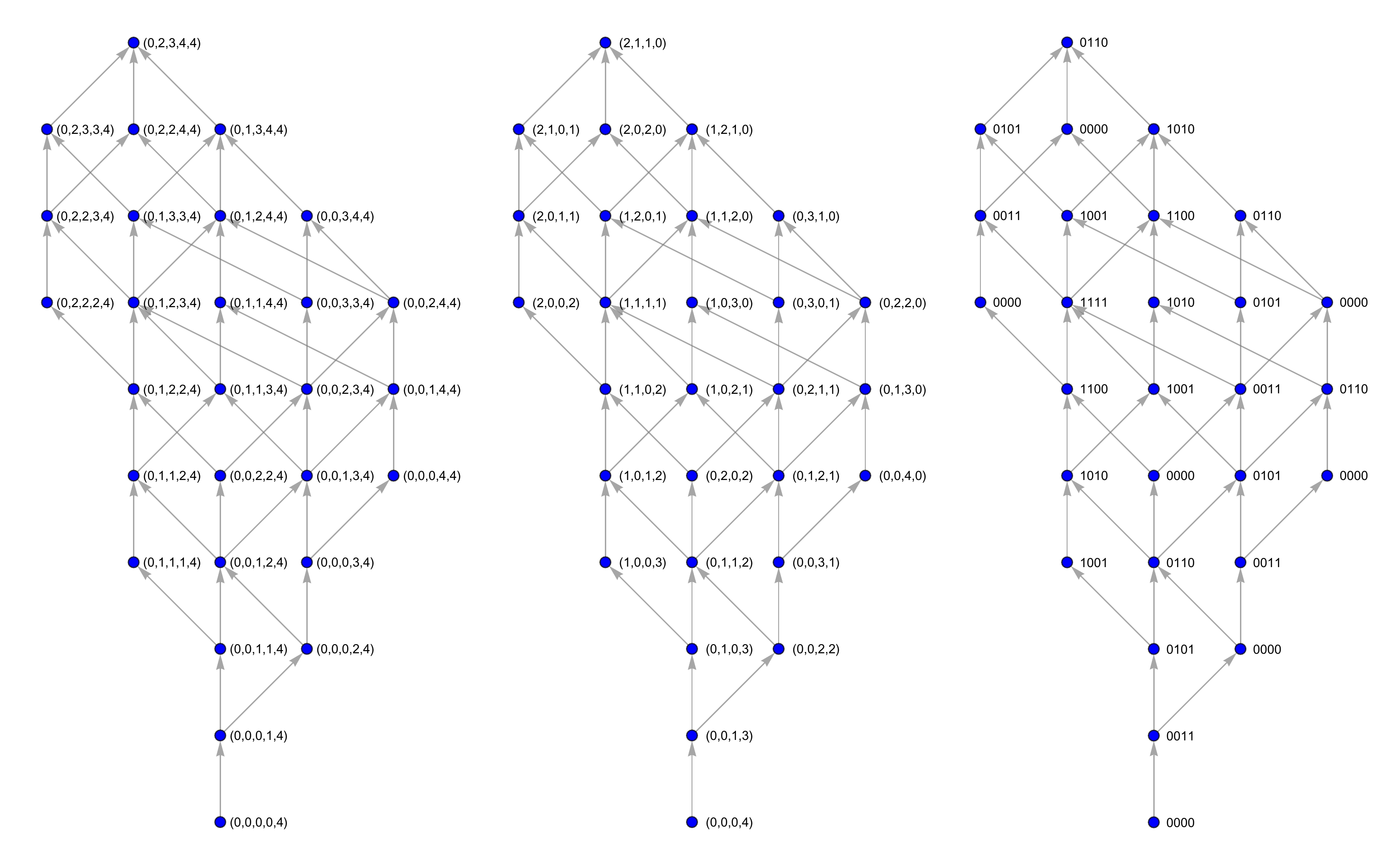}}
  \caption{For the illustration purposes we plot a Hasse diagram of $Y_{(2,3,4)}$ corresponding to the situation in Fig.~\ref{fig:circuit1Loop} and Fig.~\ref{fig:example1Loop} using three different vertex labels. On the left, the biggest element is the extended Ferrers diagram $\mu=(0,2,3,4,4)$. Indeed, the biggest reflected Ferrers diagram that fits the gray shaded polygon in Fig.~\ref{fig:example1Loop} corresponds to partition $(2,3,4)$. In the middle picture we depict the first differences~\eqref{eq:firstDiffs} interpreted as photon detection patterns~$\n$. On the right, we show bit strings as the action of the parity function~$\wp_0$ on~$\n$.}
  \label{fig:Hasse1}
  \end{figure}
\end{exa}
\begin{exa}
  \begin{figure}[h]
  \resizebox{8cm}{!}{\includegraphics{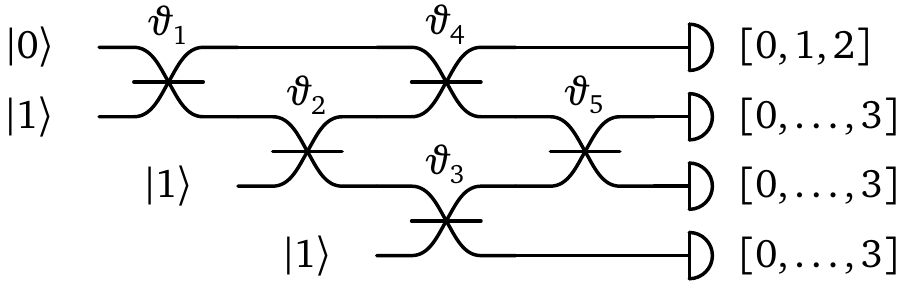}}
  \caption{Two `slices' of the triangle Reck decomposition of $\mathrm{U}(4)$  depicting a deeper circuit compared to Fig.~\ref{fig:circuit1Loop}. }
  \label{fig:circuit2Loops}
  \end{figure}
  \begin{figure}[h]
  \resizebox{13cm}{!}{\includegraphics{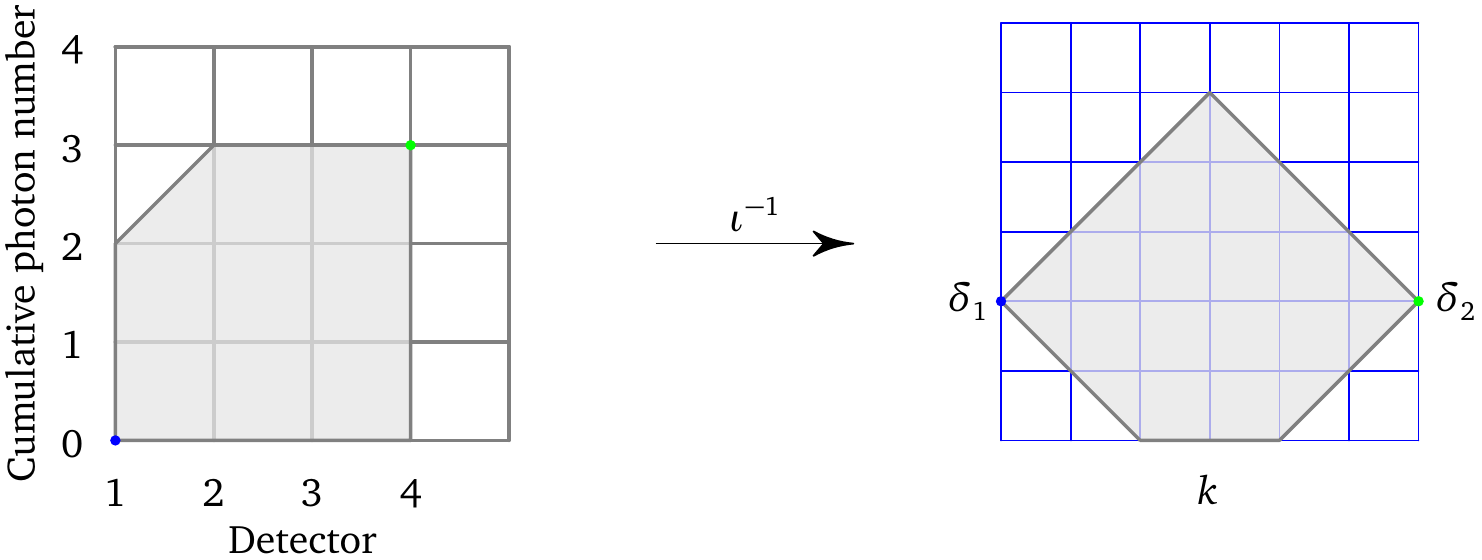}}
  \caption{The shaded polygon in the left plot captures the possible detection patterns of the circuit in~Fig.~\ref{fig:circuit2Loops}. The right plot is the Dyck path isomorphism and we can read off the parameters $k=6,\d_1=\d_2=2$ in order to count the number of detection patterns by~\eqref{eq:CardDyckPaths}.}
  \label{fig:example2Loops}
  \end{figure}
  Our second example is in Fig.~\ref{fig:circuit2Loops} illustrating the fact that we can use our analysis for any optical circuit, including the deep ones. The corresponding lattice diagrams are in Fig.~\ref{fig:example2Loops} and the description from the previous example carries over. Here we take the opportunity to illustrate a Catalan subspace inclusion mentioned earlier. Looking at Fig.~\ref{fig:reck} showing Reck's scheme for four modes, we can see that a natural slicing of $U\in\mathrm{U}(4)$ is $U=U_3U_2U_1$, where $U_1$ consists of beam-splitters 1,2 and 3 (labeled by the angle $\vt_i$), $U_2$ is the action of beam-splitters 4 and 5 and $U_3$ is beam-splitter number six. The first slice, $U_1$, will generate a Catalan Hilbert space $\rC_1(4,3)$, the first two slices, $U_2U_1$, will generate $\rC_2(4,3)$ and the whole circuit $U$ corresponds to $\rH^+(4,3)$. Hence, we get
  \begin{equation}\label{eq:exmplInclusion}
    \rC_1(4,3)\subsetneq\rC_2(4,3)\subsetneq\rH^+(4,3)
  \end{equation}
  depicted in Fig.~\ref{fig:Hasse2}.
  \begin{figure}[h]
  \resizebox{11cm}{!}{\includegraphics{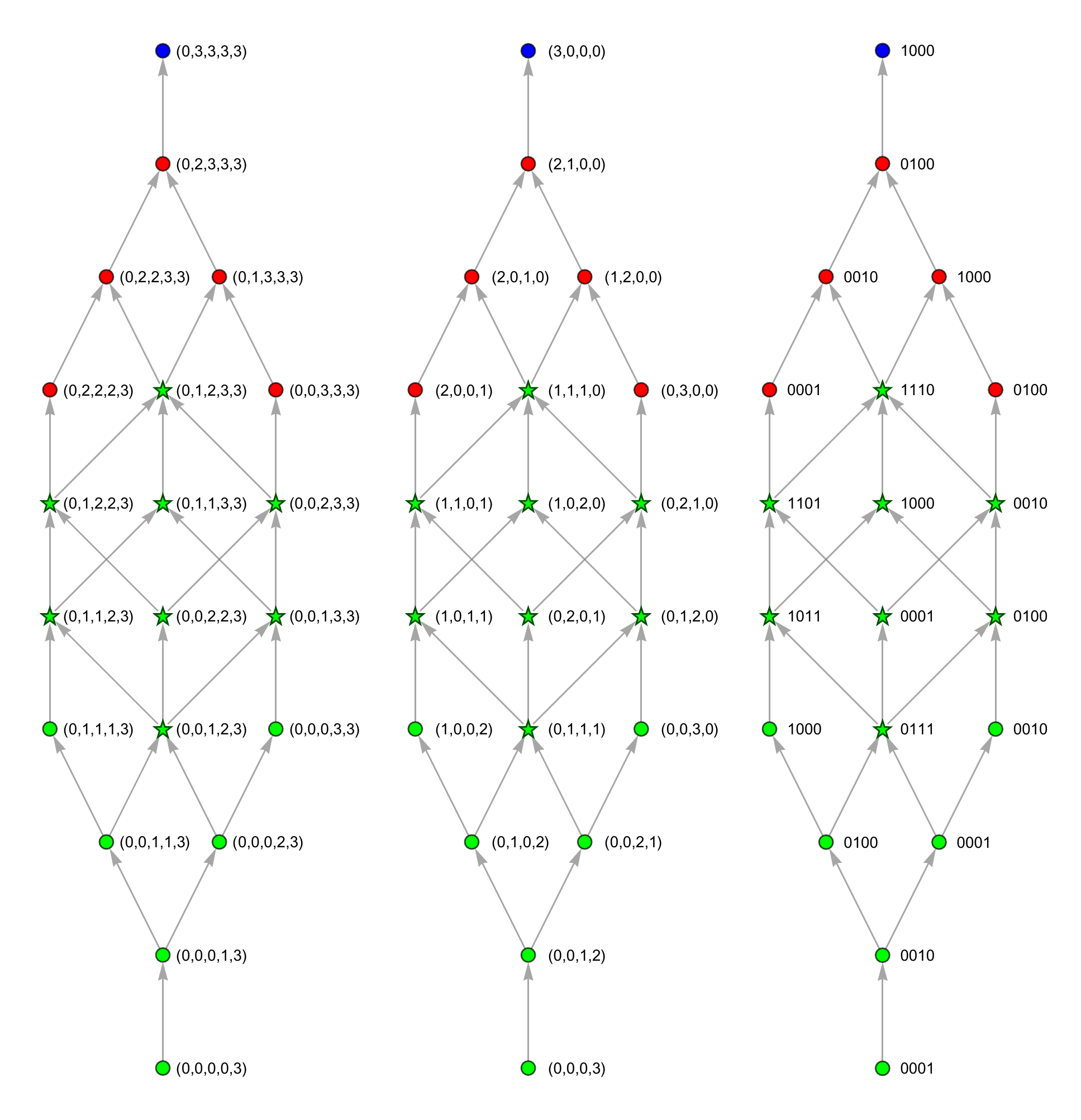}}
  \caption{Inclusion~\eqref{eq:exmplInclusion} illustrated. The Hasse diagram (blue, red and green vertices) of $Y_{(3,3,3)}$ corresponds to $\rH^+(4,3)$. The subdiagram on red and green vertices is $\rC_2(4,3)$ and the green vertices correspond to $\rC_1(4,3)$. The difference between green dots and stars is to emphasize the presence of the Boolean lattice $\rB_3$ as discussed in Sec.~\ref{s:details}.}
  \label{fig:Hasse2}
  \end{figure}
  The Catalan Hilbert space dimensions can be found from~Fig.~\ref{fig:example2Loops} and its slight modifications. The middle Hilbert space $\rC_2(4,3)$ is depicted  and we find $|\rC_2(4,3)|=|\euD(6,2,2)|=19$. For $\rC_1(4,3)$ the diagonal line of the polygon on the left starts at $(1,1)$ and ends at $(3,3)$. The mapping $\iota^{-1}$ reveals $k=6,\d_1=\d_2=1$  so  $|\rC_1(4,3)|=|\euD(6,1,1)|=14$ which is a Catalan number. Finally, the biggest space $\rH^+(4,3)$ would be an entire gray square on the left being mapped by $\iota^{-1}$ to a diamond confining all Dyck paths with the parameters $k=6,\d_1=\d_2=k/2=3$. We get $|\rH^+(4,3)|=\binom{k}{k/2}=20$ from~\eqref{eq:CardDyckPaths} in accordance with the standard bosonic formula $\binom{n+M-1}{n}$ for $n=M-1$ and $M=k/2+1$. We can also verify the results by counting the vertices in Fig.~\ref{fig:Hasse2}.
\end{exa}
In Sec.~\ref{s:parityFcn} we investigated the largest Catalan Hilbert space in the sense of~\eqref{eq:inclusionCatalan}, $\rH^+(M,n)$ -- the completely symmetric space of $n$ photons in $M$ modes, and proved some interesting properties of the parity function $\wp_j$. But this space is not something accessible in a lab as $M$ grows. The depth of the circuit makes the output heavily contaminated by errors, namely photon loss. Here we will focus on the smallest Catalan Hilbert space $\rC_1(M,n)$ as the simplest space and the simplest experimental setup when $n=M,M-1$ and at most one photon per mode, see Sec.~\ref{s:TBI}. Crucially, however,  whatever we prove for $\rC_1(M,n)$ will hold for any $\rC_i(M,n)$ thanks to~\eqref{eq:inclusionCatalan}.

Our first observation is straightforward. Setting $n=M-1$, the size of $\rC_1(M,M-1)$ grows with $M$ almost as fast as the size of $\rH^+(M,M-1)$:
\begin{equation}\label{eq:HSratio}
  {|\rC_1(M,M-1)|\over|\rH^+(M,M-1)|}={{1\over1+M}{(2M)!\over M!M!}\over{(2(M-1))!\over(M-1)!(M-1)!}}={2\over M}{2M-1\over1+M}.
\end{equation}
Just like $\rH^+(M,M-1)$, even the smallest space is much larger than $\H_{M-1}$. For any $\rC_1(M,M-1)$ we get $\d_i=1$ and $k=2(M-1)$. Plugging this to~\eqref{eq:CardDyckPaths} and using~\eqref{eq:Catalan} we find
\begin{equation}\label{eq:ratioCat}
  {|\rC_1(M,M-1)|\over|\H_{M-1}|}={C_{M-1}\over 2^{M-1}}\gg1
\end{equation}
as $M$ grows. This is good news but nowhere near close in detail to what we proved for $\rH^+(M,n)$ in Sec.~\ref{s:parityFcn}, where we showed surjection of $\wp_j$ with an exact counting. It is not obvious at all that $\wp_j$ maps the smallest Catalan Hilbert space onto a set of all qubit Hilbert space basis states. But we will show that it is true and therefore it holds for all $\rC_i(M,M-1)$ thanks to~\eqref{eq:inclusionCatalan}. In particular, we will show that the parity function~$\wp_j$ maps $\rC_1(M,M-1)$ to the same qubit Hilbert space as $\rH^+(M,M-1)$. Unlike for $\rH^+(M,M-1)$ we won't provide the exact counting of how many bosonic basis states are mapped to a given qubit basis.

\begin{figure}[h]
  \resizebox{13cm}{!}{\includegraphics{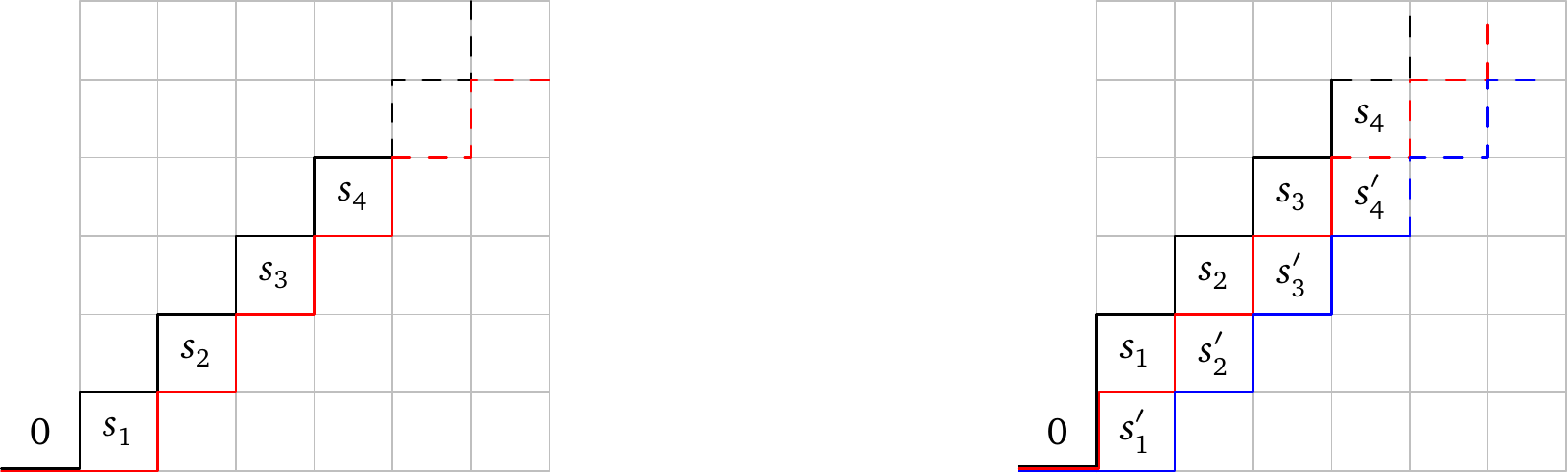}}
  \caption{On the left, we depict two extended Ferrers diagrams for $\rC_1(M,M-1)$, $\ol\la$~as the black stairs and $\ul\la$ as the red stairs, and the box bit string~$S$ introduced in~\eqref{eq:BoxBit}. All Ferrers diagram between  $\ol\la$ and  $\ul\la$ can be parametrized by~$S$. On the right, we depict the same situation for $\rC_1(M,M)$ with three Ferrers diagrams ($\ol\la$ black, $\la$ red and $\ul\la$ blue) and two box bit strings $S_1=(0,s_1,s_2\dots)$ and $S_2=(0,s'_1,s'_2\dots)$ of the same length, see Sec.~\ref{s:details}.}
  \label{fig:F4}
\end{figure}

Given $M$, let's study all extended Ferrers diagrams `between' $\ol\la=(0,1,\dots,M-1,M-1)$ and $\ul\la\df\ol\la-s=(0,0,1,2,\dots,M-1)$, where\footnote{The diagram $s$ is called a~\emph{skew} Ferrers or Young diagram and it is typically defined by the difference of the top and bottom Ferrers diagrams written as~$s=\ol\la/\ul\la$.}
\begin{equation}\label{eq:Bbit}
  s=(0,\underbrace{1,\dots,1}_{M-1},0).
\end{equation}
By between we mean any Ferrers diagram that can be obtained from $\ol\la-S$, where
\begin{equation}\label{eq:BoxBit}
  S\df(0,s_1,\dots,s_{M-1},0)
\end{equation}
and $s_i=\{0,1\}$. The sequence $S$ will be called a \emph{box bit string}. The name becomes clear if we plot $\ol\la$ and $\ul\la$ as in the left panel of Fig.~\ref{fig:F4}.
We can see that whenever $s_i=1$, we remove the corresponding box to obtain $\la$ from~$\ol\la$. From the properties of the Ferrers diagrams we can see that all $2^{M-1}$ box bit string configurations are allowed. Hence, any Ferrers diagram between $\ol\la$ and $\ul\la$ can be written as $\la=\ol\la-S$. Following~\eqref{eq:firstDiffs}, we take the first differences to get $n_i\in\n$:
\begin{equation}\label{eq:1stDiffsBoxBit}
  n_i=\la_{i}-\la_{i-1}=\ol\la_{i}-s_{i}-(\ol\la_{i-1}-s_{i-1})=\ol\la_{i}-\ol\la_{i-1}+(-s_{i}+s_{i-1}).
\end{equation}
From the form of $\ol\la$ we find the $M$-tuple $(\ol\la_{i}-\ol\la_{i-1})_1^{M}=(1,1,\dots,1,0)$ and we denote the bit differences
$$
d_i=s_{i-1}-s_{i}\in\{-1,0,1\}.
$$
The box bit strings~$S$ are all different by construction. Does it mean that  the $M$-tuples $D=(d_i)_1^{M}$ are mutually different as well? This is equivalent to asking if we can deduce the correct box bit strings $S$ from~$D$. We can because
\begin{align}\label{eqs:dOptions}
  d_i=-1 & \Rightarrow (s_{i-1},s_{i})=(0,1),\nn \\
  d_i=0 & \Rightarrow (s_{i-1},s_{i})=(0,0)\mbox{ or }(1,1),\\
  d_i=1 & \Rightarrow (s_{i-1},s_{i})=(1,0).\nn
\end{align}
The ambiguity for $d_i=0$ can be resolved by the `boundary condition' of~$S$ in~\eqref{eq:BoxBit}, namely that $s_0=s_{M}=0$. Hence, all $D$s are different. But we need to prove something stronger, namely that the action of the parity function, $\wp_j$ (sufficient for $j=0$), on~$\n$ still outputs $2^M$ different bit strings. Using~\eqref{eq:1stDiffsBoxBit} and the properties of the modular addition it acts as
\begin{equation}\label{eq:wponni}
  \wp_0(n_i)=\wp_0(\ol\la_{i}-\ol\la_{i-1})\oplus\wp_0((-s_{i}+s_{i-1}))
  =(\ol\la_{i}-\ol\la_{i-1})\oplus\wp_0(d_i).
\end{equation}
Since $\wp_0(d_i)=\wp_0(-d_i)$ we need to make sure that no first difference of $S$ is the (ordinary) negation of another one. To this end, we realize that this precisely happens in the set of all $2^{M-1}$ bit strings of length~$M-1$. Trivially, when $0\neq d_i=s_{i-1}-s_i$ then $-d_i=s_{i}-s_{i-1}$ by swapping the bits so the bit strings with all opposite bit values are the negation of each other. But the situation changes once we prepend a constant bit like the zero bit in the box bit string defined in~\eqref{eq:BoxBit}. If we split the $2^{M-1}$ bit strings in one half starting by zero and the second half starting by one (note that they necessarily contain the swaps of the first half), then prepend a zero bit and finally take the first difference then all the first half differences will start with bit zero whereas the second half differences will start with bit one, effectively flagging the swaps that would otherwise be indistinguishable under the action of $\wp_0$. By further appending a zero bit, again like in~\eqref{eq:BoxBit}, taking the first differences won't change the first flag bit and so we get $2^{M-1}$ different sequences~$D$ that are mapped by $\wp_0$ to $2^{M-1}$ different bit strings. This concludes the argument.

We have to tie up a few loose ends and add some comments.
\subsubsection{Further details}\label{s:details}
We don't have to investigate the details of actually taking the first differences and the modular sum with $(\ol\la_{i}-\ol\la_{i-1})$ in the RHS of~\eqref{eq:wponni} to find what bit strings we obtained. This is because we showed in Sec.~\ref{s:parityFcn} that for $\rH^+(M,M-1)$ the range of the parity function $\wp_0$ is $\H_{M-1}$. Because of~\eqref{eq:inclusionCatalan} it must be the same bit strings and therefore the same Hilbert space basis. We can see it explicitly in Fig.~\ref{fig:Hasse2} for $M=4$. Looking at the Hasse diagram on the right with the bit values we see that the Hilbert space basis of $\rC_1(4,3)$ (the green stars and points) are the same ones as in the whole structure (red+green+blue vertices) corresponding to $\rH^+(4,3)$ (cf.~\eqref{eq:exmplInclusion}).
\begin{figure}[t]
  \resizebox{8cm}{!}{\includegraphics{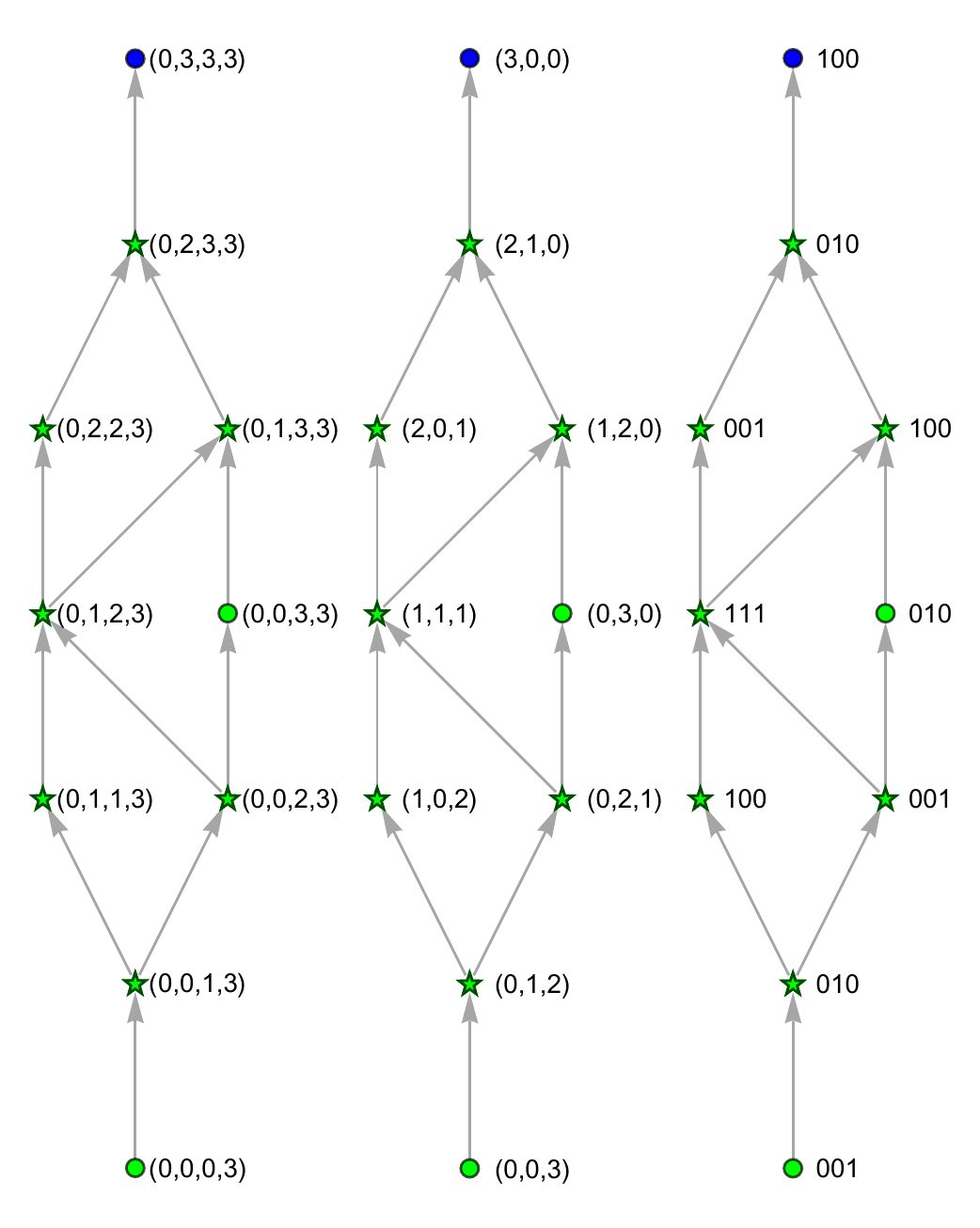}}
  \caption{The Hasse diagrams of $Y_{(3,3)}$ for $\rH^+(3,3)$. The Catalan Hilbert space $\rC_1(3,3)$ corresponds to the green subdiagram and the green stars are the ordinal sum of two Boolean lattices~$\rB_2\boxplus\rB_2$.}
  \label{fig:Hasse3}
\end{figure}

We argued in Sec.~\ref{s:parityFcn} that for a given $M$, to access a bigger qubit Hilbert space $\H_M$, we run two experiments, one for $n=M-1$ and the other one for $n=M$. Let's focus on the latter, again for the smallest Catalan Hilbert space $\rC_1(M,M)$ in~\eqref{eq:inclusionCatalan}. Compared to the left plot in Fig.~\ref{fig:F4}, the generic highest Ferrers diagram is shifted up by one box, see $\ol\la$ in the right plot of Fig.~\ref{fig:F4}. By the same procedure we employed for $n=M-1$ we can therefore introduce two box bit strings~\eqref{eq:BoxBit} of the same length, $S_1,S_2$, one for each layer of boxes we intend to remove. The procedure goes exactly in the same way as described above except for being performed twice. The top extended Ferrers diagram $\ol\la=(0,2,3,\dots,M,M)$ becomes the middle one $\la=\ol\la-S_1=(0,1,2,\dots,M-1,M-1)$ for $S_1=(0,1,\dots,1,0)$, just to serve as a starting point for the second round ending with the bottom Ferrers diagram $\ul\la=\la-S_2=(0,0,1,2,\dots,M-1)$ for $S_2=(0,1,\dots,1,0)$. We illustrate the case of $\rC(3,3)$ in Fig.~\ref{fig:Hasse3}.

\begin{figure}[h]
  \resizebox{12cm}{!}{\includegraphics{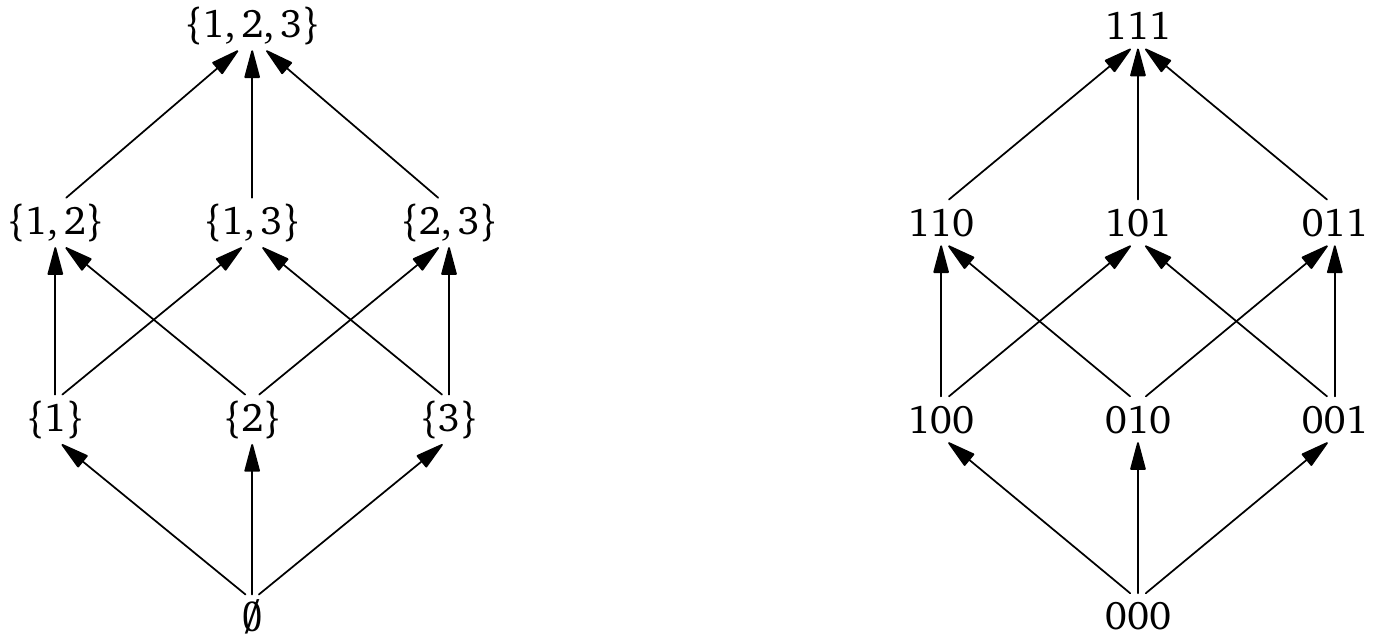}}
  \caption{(Left) The Boolean lattice $\rB_3$  of all subsets of the power set of $Z=[1,2,3]$ ordered by inclusion. (Right) The isomorphic lattice, where the three-bit vertices $p=p_1p_2p_3$ index the elements of $Z$. The complement  of $p_i$ is $s_i$ from $S$ in~\eqref{eq:BoxBit} used to track the removed boxes from $\ol\la$. For example, $p=110$ corresponds to $S=(0,0,0,1,0)$ telling us to remove one box labeled $s_3$, see Fig.~\ref{fig:F4}. This is how the finite Boolean lattices make their appearance in Young's sublattices we study.}
  \label{fig:B3iso}
\end{figure}

As previously discussed, the Hasse diagrams we study are special posets called Young's (sub)lattices~$Y_\mu$. In fact, our procedure of removing the upper layer of boxes from $\ol\la$ using the box bit string emphasizes even more fundamental structure contained in every Young's lattice: the \emph{finite Boolean lattice}~$\rB_n$ -- a truly fundamental object in lattice theory~\cite{birkhoff1940lattice,stanley1999enumerative}. For a given finite set $Z$, where we denote $n=|Z|$, we may form a power set~$P(Z)$ (the set of all subsets) of cardinality $2^{n}$. If we add the operations of set union (in the role of the join operation) and intersection (as the meet) then the power set is isomorphic to the Boolean lattice $\rB_n=(P(Z),\,\subseteq)$. How does the Boolean lattice materialize in our case? Consider the set $Z=[M-1]$ and take the binary negation of the substring $p\df(\neg s_1,\dots,\neg s_{M-1})$ of the box bit string~$S$ in~\eqref{eq:BoxBit}. Clearly, $|p|=2^{M-1}$ which is equal to the power set cardinality of~$Z$. We index the elements of the power set $P$ by picking the $i$-th element of~$Z$ iff $\neg s_i=1$. For example, if $\neg s_i=0,\forall i$ we obtain the empty set $\emptyset$. We exemplified the isometry in Fig.~\ref{fig:B3iso} for $M=4$. We indeed recognize the similar projected `cubes' of the Boolean lattice $\rB_3$ in Fig.~\ref{fig:Hasse2} emphasized by the green star vertices.
The graph of the Boolean lattice~$\rB_n$ is isomorphic to the hypercube graph~$H_n$ but, in general, we cannot rely on this visual aid when asking how many Boolean sublattices a lattice has. For example, there is one projected cube visible in the green Hasse subdiagram of Fig.~\ref{fig:Hasse2} and this is indeed the only $\rB_3$. However, despite counting 9 `squares' corresponding to $\rB_2$ there are in fact in total 21 (green) Boolean lattices~$\rB_2$. Similarly in Fig.~\ref{fig:Hasse3}, there are 3 `visible' $\rB_2$'s but also two less visible (that is, not visually aided by the edges).


We can now review the structure of $\rC_1(3,3)$, where we introduced two box bit strings. Indeed, we can see two hypercube graphs $H_2$ (squares) sharing one vertex corresponding to the middle Ferrers diagram. In summary, by showing that $\wp_0$ is a surjection we also refined the inclusion diagram~\eqref{eq:inclusionCatalan} to become
\begin{equation}\label{eq:inclusionCatRef}
\rB_{M-1}\subsetneq\rC_1(M,n)\subsetneq\cdots\subsetneq\rC_{M-2}(M,n)\subsetneq\rH^+(M,n).
\end{equation}
We are abusing the notation a bit since $\rC_i(M,n)$ is defined as a Hilbert space but we now see it as Young's sublattice whose vertices are the spanning basis of the said Hilbert space. What is our ultimate goal here? It is to elucidate the structure of $\rC_i(M,n)$ and we are getting a glimpse of it by the presence of the Boolean lattice. On the practical side we would like to know how exactly the parity function $\wp_0$ maps the set of $\rC_i(M,n)$ basis state to that of $\H_{M-1}$ (or $\H_M$ for $\wp_j$).

We can't answer the second question here but we are able to say something more. Note that after reaching the lower Ferrers diagram $\ul\la$ using the box bit string $S$~in~\eqref{eq:BoxBit} we can introduce a one bit shorter bit boxstring and repeat the procedure with $\ul\la$ starring as the new top Ferrers diagram. We again encounter a (smaller) Boolean lattice all the way to the smallest one, $\rB_1$, thus further refining~\eqref{eq:inclusionCatRef}. Indeed, we recognize it in all our previous examples, for instance in Fig.~\ref{fig:Hasse2}, we see $\rB_2$ as the square `hanging' from the previously identified Boolean lattice~$\rB_3$ and at the very bottom the two-element algebra $\rB_1$, sharing one element with the $\rB_2$. The process  where the smallest element of a lattice $\rL_2$ is identified with the highest element of another lattice~$\rL_2$ is called the \emph{ordinal sum of lattices} $\rL_2\boxplus\rL_1$ which is clearly not commutative~\cite{deiters2009sums}\footnote{Cf. with the \emph{vertical sum of posets} as its generalization, see~\cite{davey2002introduction}.}. So our first refinement of~\eqref{eq:inclusionCatalan} is
\begin{equation}\label{eq:inclusionCatRefRef}
\bigboxplus_{i=M-1}^{1}m_i\rB_{i}\subsetneq\rC_1(M,n)\subsetneq\cdots\subsetneq\rC_{M-2}(M,n)\subsetneq\rH^+(M,n),
\end{equation}
where $m_i\geq1$ is the multiplicity of $\rB_i$. An example of $\underbrace{\rB_2\boxplus\rB_2}_{2\rB_2}\boxplus\,\rB_1$ is in Fig.~\ref{fig:Hasse3} ($2\rB_2$ are the green stars and $\rB_1$ is the two bottom vertices sharing an edge). It follows from the combinatorial argument involving the total number of Ferrers diagrams, cf.~\eqref{eq:ratioCat}, that the first inclusion of~\eqref{eq:inclusionCatRefRef} becomes extremely sharp as $M$ grows. This is not satisfactory.

Generalizing our prescription from Fig.~\ref{fig:F4}, where we found the biggest Boolean lattice~$\rB_M$, we can count the multiplicity of the biggest (visible) Boolean lattices. Indeed, whenever we are able to remove $M$ boxes from a Ferrers diagram of $M$ columns such that the resulting diagram is a valid Ferrers diagram we encountered $\rB_M$. For the Ferrers diagram $\mu=(1,2,\dots,M)$ there is only one $\rB_M$, like in the left picture of Fig.~\ref{fig:F4}. But, on the right for $\mu=(2,3,\dots,M)$, there are plenty of possibilities. As an example, consider $\mu=[2,3,4]$ from Fig.~\ref{fig:Hasse1}. We can remove three boxes in four different ways (in a certain order). Looking at the figure, there are hiding four $\rB_3$ Boolean lattices. Since they are maximal, we know what qubit bases their vertices are  decorated with (in the right panel). The problem is, however, that they are not disjoint. Moreover, smaller Boolean sublattices are present too and their counting seems unruly. Perhaps more insight would be obtained by a \emph{chain decomposition} of  Young's lattice. A chain is a totally ordered subset (i.e., all its elements are mutually comparable) of a poset. A chain decomposition of lattices is, however, a notoriously difficult problem in lattice theory. We leave this question open.

\section*{Acknowledgement}
The authors would like to thank Josh Nunn, Richard Murray, William Clements, Kris Kaczmarek and Adel Sohbi for discussions and comments and especially Alex Neville for sharing his code helping us simulate the output statistics of a boson sampling device and Richard Tatham for carefully reading the manuscript.

\section*{Appendix}
\subsection*{Parameters of some simulations from the main text}
A random QUBO matrix used in Sec.~\ref{s:apps}:
\begin{equation}
	Q =
	\left[\begin{smallmatrix}
		-0.128& -0.445& -0.022& -0.082& -0.012& -0.642& -0.439& -0.368&
		0.273&  0.148& -0.115\\
		-0.445& -0.731& -0.39 & -0.31 &  0.213&  0.101& -0.46 &  0.644&
		-0.808&  0.283& -0.021\\
		-0.022& -0.39 & -0.746& -0.016& -0.337& -0.826&  0.203& -0.381&
		-0.219& -0.664&  0.06 \\
		-0.082& -0.31 & -0.016&  0.587&  0.357&  0.156&  0.142&  0.547&
		-0.458&  0.758& -0.118\\
		-0.012&  0.213& -0.337&  0.357&  0.071&  0.924&  0.048& -0.892&
		0.105&  0.6  &  0.015\\
		-0.642&  0.101& -0.826&  0.156&  0.924&  0.601& -0.075&  0.427&
		-0.173&  0.092&  0.35 \\
		-0.439& -0.46 &  0.203&  0.142&  0.048& -0.075& -0.481& -0.226&
		0.047&  0.28 &  0.003\\
		-0.368&  0.644& -0.381&  0.547& -0.892&  0.427& -0.226& -0.006&
		-0.168& -0.428& -0.272\\
		0.273& -0.808& -0.219& -0.458&  0.105& -0.173&  0.047& -0.168&
		0.288&  0.58 & -0.238\\
		0.148&  0.283& -0.664&  0.758&  0.6  &  0.092&  0.28 & -0.428&
		0.58 &  0.184& -0.164\\
		-0.115& -0.021&  0.06 & -0.118&  0.015&  0.35 &  0.003& -0.272&
		-0.238& -0.164&  0.096
	\end{smallmatrix}\right].
\end{equation}
A random QUBO matrix used in Sec.~\ref{s:deeperCircuit}:
\begin{equation}
	Q =
	\left[\begin{smallmatrix}
		-0.1280102 & -0.76942513 & -0.31575758 &  0.28188388 & -0.45247223 &
		-0.20187769\\
		-0.76942513&  0.23854193& -0.18676721& -0.65352725&  0.21787914&
		-0.26911468\\
		-0.31575758& -0.18676721& -0.63112027&  0.29058124&  0.07998729&
		0.13464356 \\
		0.28188388& -0.65352725&  0.29058124& -0.86942699& -0.46493199&
		-0.42039925\\
		-0.45247223&  0.21787914&  0.07998729& -0.46493199& -0.55938759&
		-0.14493699\\
		-0.20187769& -0.26911468&  0.13464356& -0.42039925& -0.14493699&
		-0.2262147
	\end{smallmatrix}\right].
\end{equation}
Learning curves for the results presented in Fig.~\ref{fig:2loops}.
\begin{figure}[h]
	\begin{subfigure}[t]{7cm}
		\resizebox{7cm}{!}{\includegraphics{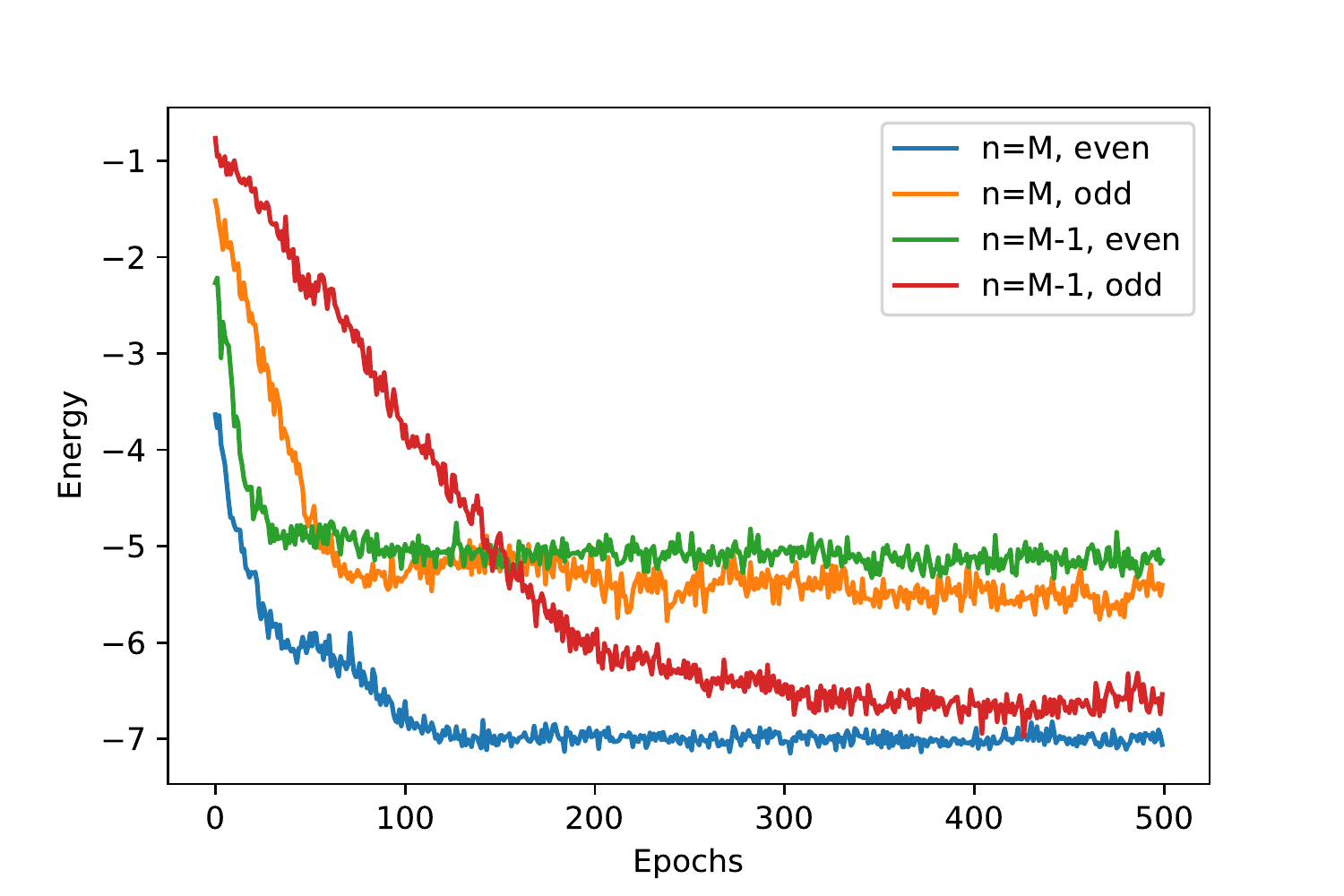}}
	\end{subfigure}
	\begin{subfigure}[t]{7cm}
		\resizebox{7cm}{!}{\includegraphics{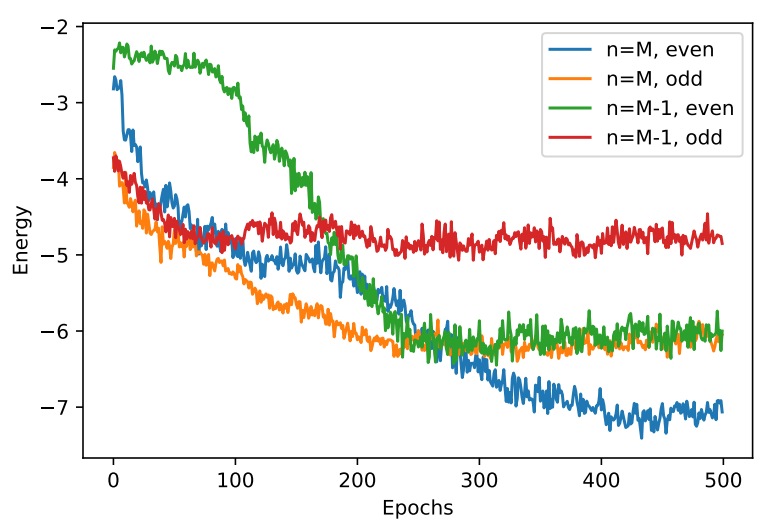}}
	\end{subfigure}
	\caption{Additional graphs accompanying Fig.~\ref{fig:2loops}. (Left) Learning curves the shallowest circuit addressing Hilbert space~$\rC_1(6,n)$ for $n=5,6$. (Right) Learning curves for a deeper circuit~$\rC_2(6,n)$. In the deeper circuit case, the system takes more time to learn as expected thanks to an increased number of parameters.}
	\label{fig:2loopsAnnexe}
\end{figure}
The three smallest energies this problem are $-7.92, -7.30$ and $-5.89$.

\bibliographystyle{unsrt}


\end{document}